\documentclass[12pt]{article}
\usepackage{amsmath,geometry}
\usepackage{fullpage}
\usepackage{natbib}
\usepackage{algorithm}
\usepackage{algpseudocode}
\usepackage{lscape}
\usepackage{setspace}
\usepackage{authblk}
\usepackage{graphics,graphicx}
\usepackage{appendix}
\usepackage{subfig}
\usepackage{url}
\usepackage{amsmath,amssymb,amsthm}
\usepackage{bm}
\usepackage{graphicx}
\usepackage{booktabs}
\usepackage{comment}
\usepackage{booktabs}
\usepackage{varwidth}
\newsavebox\tmpbox

\newcommand{\betavec}{\boldsymbol{\beta}}

\newcommand{\zerovec}{\boldsymbol{0}}
\newcommand{\onevec}{\boldsymbol{1}}

\newcommand{\evec}{\boldsymbol{e}}

\newcommand{\yvec}{\boldsymbol{y}}
\newcommand{\zvec}{\boldsymbol{z}}

\newcommand{\I}{\textbf{I}}

\newcommand{\Smat}{\textbf{S}}

\newcommand{\X}{\textbf{X}}

\DeclareMathOperator*{\argmin}{arg\,min}

\begin{document}

\title{A Graphical Comparison of Screening Designs using Support Recovery Probabilities}

\author[1]{\small Kade Young}
\author[2]{\small Maria L. Weese}
\author[1]{\small Jonathan W. Stallrich}
\author[3]{\small Byran J. Smucker}
\author[4]{\small David J. Edwards}
\affil[1]{\small Department of Statistics, North Carolina State University, Raleigh, NC}
\affil[2]{\small Department of Information Systems \& Analytics, Miami University, Oxford, OH}
\affil[3]{\small Department of Statistics, Miami University, Oxford, OH}
\affil[4]{\small Department of Statistical Sciences and Operations Research, Virginia Commonwealth University, Richmond, VA}

\date{}


\maketitle
\doublespace

\begin{abstract}

A screening experiment attempts to identify a subset of important effects using a relatively small number of experimental runs. Given the limited run size and a large number of possible effects, penalized regression is a popular tool used to analyze screening designs. In particular, an automated implementation of the Gauss-Dantzig selector has been widely recommended to compare screening design construction methods. Here, we illustrate  potential reproducibility issues that arise when comparing screening designs via simulation, and recommend a graphical method, based on screening probabilities, which compares designs by evaluating them along the penalized regression solution path. 
This method can be implemented using simulation, or, in the case of lasso, by using exact local lasso sign recovery probabilities. Our approach circumvents the need to specify tuning parameters associated with regularization methods, leading to more reliable design comparisons.  This article contains supplementary materials including code to implement the proposed methods.


\end{abstract}

\noindent%

\section{Introduction}\label{sec:Introduction}

A screening design, ideally utilized as the first step in a sequential experimentation process, attempts to select which of $k$ factors truly drive a response, using a relatively small number of runs, $n$. For screening scenarios, the
linear model of interest is 
\begin{align}\label{eq:me_model}
    \yvec=\onevec\beta_0+\X\betavec+\evec,
\end{align}
\noindent where $\evec \sim N(\zerovec,\sigma^2\I)$, $\X$ the $n \times p$ model matrix, and $\betavec$ a $p$-vector of parameters. The $p$ columns in the model matrix may be comprised solely of the $k$ main effects, or may include some or all of the $k(k-1)/2$ two-factor interactions. The $k$ main effect columns of $\X$ represent the experimental design with each row providing the settings of the $k$ factors for a given run. For simplicity, in the examples in this work we consider only two-level factors taking on values $\pm 1$. The analysis goal is to uncover the subset of these factorial effects that are \textit{active}; that is, those effects with non-trivially non-zero parameter values. We define the \textit{active set} $\mathcal{A}$ as the set of indices of the active effect columns.  To maximize the chance that the experiment is a success, we should identify the design and analysis method that is best at estimating $\mathcal{A}$.

We will focus on designs where $n < p+1$ so that a unique least squares estimator for $\betavec$ does not exist. It is possible to use an approach like forward selection, but it has notable shortcomings \citep{westfall1998forward} and performs poorly compared to other methods \citep{MarleyWoods10, draguljic_etal2014, weese_etal2015}. Instead, regularization methods are often recommended as a general purpose analysis approach, for which estimation is done under a penalty on the magnitude of the elements of $\betavec$, shrinking them towards zero. The amount of shrinkage is controlled by a tuning parameter, $\lambda>0$, with larger $\lambda$ enforcing greater shrinkage. 
It is common practice to compare the ability of designs to estimate $\mathcal{A}$ via simulation studies under one or more regularization analysis methods.

This paper demonstrates the sensitivity of the simulation-based design comparisons to different sets of $\lambda$'s and tuning parameter selection methods. As an alternative, we recommend a robust graphical approach, motivated by \citet{stallrich2023optimal}, which compares designs under a regularization method by plotting the probability of perfect estimation of $\mathcal{A}$ as a function of $\lambda$. We will focus our examples and discussion on two types of screening experiments. The first is supersaturated designs where $n \leq k$ and an analysis that assumes no active two-factor interactions. The second type is screening experiments with $k+1 \leq n < p$ that assume some active two-factor interactions.


\subsection{The Gauss-Dantzig Selector}

Various analysis methods have been suggested for analyzing screening designs, including the Dantzig selector \citep{candes_tao2007}, lasso \citep{tibshirani1996regression}, and other regularization or Bayesian methods. The Dantzig selector has been the most omnipresent. This is perhaps due to its early introduction to the SSD analysis literature by \citet{phoa2009analysis}. The Dantzig selector can be defined as follows. After centering $\yvec$ to have mean 0 and centering and scaling the columns of $\X$, estimates of $\betavec$ can be obtained as
\begin{align}
    \hat{\betavec}_{DS} = \argmin_{\betavec}||\betavec||_1 & \text{ subject to } ||\X^T(\yvec - \X\betavec)||_{\infty} \leq \lambda\ , \ \label{eq:DS}
\end{align}
where $||\cdot||_{\infty}$ denotes the maximum element in absolute value of the vector argument and $\lambda\geq 0$ is a tuning parameter. In the context of screening designs, the recommended practice is to investigate the plot of the estimates versus values of $\lambda$ \citep{phoa2009analysis, weese2021strategies}. This approach avoids the need to specify a tuning parameter selection method. For simulation studies, such as those used to compare design construction and analysis methods, viewing plots is, unfortunately, not a viable option. \cite{phoa2009analysis} suggested a multi-stage automated approach known as the Gauss-Dantzig selector (GDS). GDS involves an additional tuning parameter, $\gamma >0$, being a hard threshold applied to the estimates. \cite{weese2021strategies} describes the GDS approach as follows:

\begin{enumerate}
    \item Solve (\ref{eq:DS}) for $d$ evenly spaced values of $\lambda$ over the interval $[0,  ||\X^T\yvec||_\infty]$ (excluding the endpoints). Denote the $d$ solutions to (\ref{eq:DS}) as $\hat{\betavec}_{DS}(\lambda)$.
    \item For each $\hat{\betavec}_{DS}(\lambda)$ and some $\gamma > 0$, set all $|\hat{\beta}_j(\lambda)| < \gamma$ to $0$. Denote this thresholded solution as $\hat{\betavec}_{DS}(\lambda , \ \gamma)$.
    \item For each $\hat{\betavec}_{DS}(\lambda , \ \gamma)$, calculate the OLS estimates using only the columns of $\X$ that correspond to the nonzero $\hat{\beta}_{j}(\lambda , \ \gamma)$. Calculate the value of some information criteria (e.g., AICc or BIC). 
    \item Select the $\hat{\betavec}_{DS}(\lambda , \ \gamma)$ with the smallest information criteria value and classify the effects corresponding to nonzero $\hat{\beta}_{j}(\lambda , \ \gamma)$ as active, and the others as inactive. 
\end{enumerate}

Table \ref{tab:dant_param} lists recent papers that use the GDS to compare screening design construction methods and/or analysis methods for screening designs. GDS was implemented in all simulations in the papers listed in Table~\ref{tab:dant_param}, but the implementations of this procedure have varied substantially. Of course, the implementation of GDS depends on the particular simulation scenario and is up to author discretion; however, the choice of the GDS tuning parameters impact the analyses and thus the simulation results.  The choice of $\gamma$ has obvious consequences and is usually reported when GDS is used, but more subtle choices like the coarseness of the search grid for values of $\lambda$ (i.e., the size of $d$) are often not explicitly discussed.

            

\begin{table}[ht]
	\centering
    \footnotesize
	\label{my-label}
	\begin{tabular}{@{}llcccccccccc@{}}
		\toprule
		& Paper & $\gamma$ & Selection & $\lambda$ values ($d$) \\ \midrule
		& \cite{phoa2009analysis} & 0 or 0.25*max$|\hat{\beta_j}|$  & mAIC  & ``hundreds" \\
		& \cite{MarleyWoods10} & $\sigma$   & BIC  &   --    \\
		& \cite{draguljic_etal2014} & $\mu_{act}-3.5\sigma_{act}$   & AIC & -- \\
		& \cite{chen2013screening}   & 1 & mAIC  &  --    \\
		& \cite{phoa2013stepwise}   & 0.1*max$|\hat{\beta_j}|$  & mAIC & --  \\
        SSD    & \cite{huang2014functionally} & 1  & --  & -- \\
		$n \leq k$ & \cite{weese_etal2015}   & 1.5 & BIC  & by 1 (0 to max($\X^T\yvec$)) \\
         & \cite{drosou2019sure} & -- & mAIC & -- \\
		& \cite{weese2017powerful}    & 1.5  & BIC  & by 1 (0 to max($\X^T\yvec$)) \\ 
            & \cite{jones2020construction} & $\sigma$ & AICc & --   \\
            & \cite{weese2021strategies} & 0.1*max$|\hat{\beta_j}|$ & BIC & by 1 (0 to max($\X^T\yvec$))  \\  
            & \cite{singh2022selection} & 0.1*max$|\hat{\beta_j}|$ & BIC  & 12 steps (0 to max($\X^T\yvec$)) \\  \midrule
            & \cite{mee2017selecting} & 0.5 & AICc & by 1  (0 to max($\X^T\yvec$)) \\
            Traditional & \cite{errore2017using} & -- & AICc & -- \\
            $n \geq k+1$ & \cite{weese2018analysis} & $\sigma$  & BIC/AICc & by 1  (0 to max($\X^T\yvec$)) \\
             & \cite{vazquez2021mixed} & 0.5 & AICc & -- \\
            & \cite{hameed2022analysis} & $\sigma$ & AICc &  -- \\
            \bottomrule
	\end{tabular}
\caption{  Parameters used in the implementation of the Dantzig selector for SSDs (top row) and larger screening designs (bottom row). }
\label{tab:dant_param}
\end{table}

\begin{figure}[ht]
    \centering
    \includegraphics[width=5in]{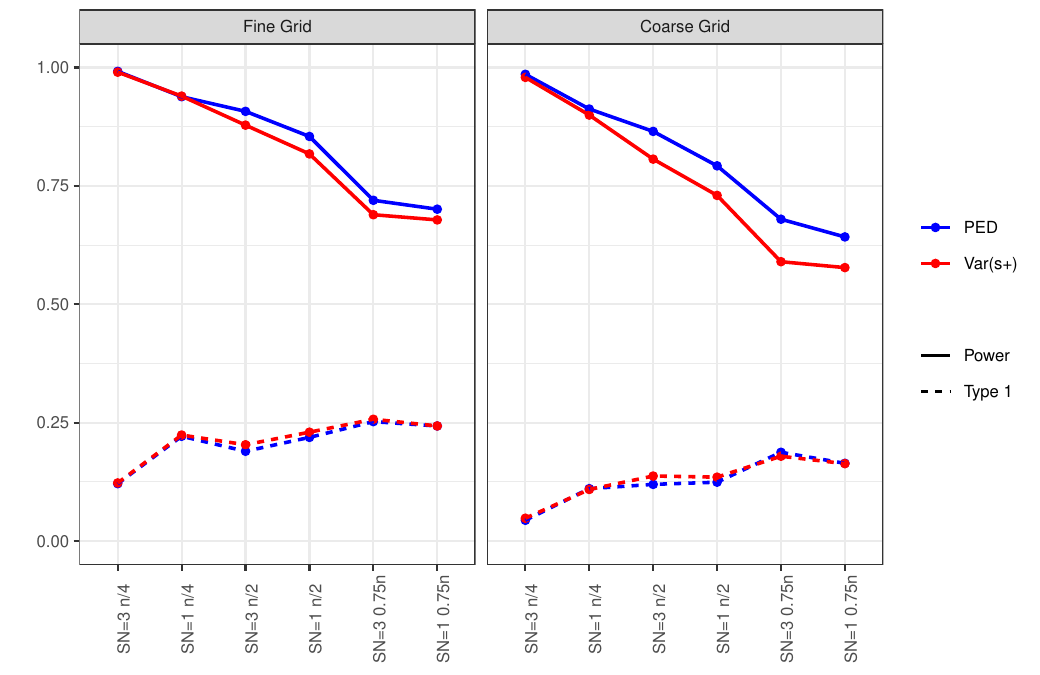}
    \caption{Power and type 1 error of $n=14$ and $k=24$ PED and Var(s+) SSDs with the Gauss-Dantzig selector implemented using a ``fine'' search grid and a ``coarse'' search grid. }
    \label{fig:coarse_fine}
\end{figure}

To briefly illustrate the impact of the choice of the tuning parameters on the results of GDS, we repeat the simulations from \cite{singh2022selection} which compared Pareto Efficient Designs (PEDs) with $Var(s+)$ SSDs for $n=14$ and $k=24$.  The signs of the truly active effects were assumed to be known and thus set as positive. Two signal-to-noise ratios (SN) were considered in which the true model effects were generated according to $Exp(1)+SN$ and errors are simulated from a standard normal distribution. The number of truly active factors range from $0.25n$ to $0.75n$. Truly inactive effects have $\beta_j=0$.  For each scenario, we randomly generated 1000 response vectors and implemented GDS with $\gamma=0.1*\max(|\hat{\beta_j}|)$ and BIC, but used two different search grids for $\lambda$.  The ``Fine Grid" includes $d=100$ values of $\lambda$ between 0 and max$(\X^T\yvec)$.  The ``Coarse Grid'' contains $d=12$ values of $\lambda$ between 0 and max$(\X^T\yvec)$ where the first and the last values were removed.  Figure~\ref{fig:coarse_fine} shows the comparisons between the two SSDs in terms of power (proportion of correctly selected active factors) and type I error (proportion of incorrectly selected active  factors).  Although the difference is subtle, different conclusions might be reached when viewing the results from the fine grid, where there appears to be little difference between the PED and $Var(s+)$ SSD and the coarse grid where the  PED design clearly has improved power. 

\subsection{GDS and Supersaturated Designs}

Analyzing supersaturated designs (SSDs) is challenging because not even the main effects can be estimated using least-squares. The majority of the SSD construction methods aim for designs as close to orthogonal as possible (i.e., $\X^T\X \approx n\I$). This ideal structure is approximated through various heuristic optimality measures, such as the $E(s^2)$ criterion \citep{Booth62}, which chooses among designs with balanced columns (i.e., $\X^T\onevec =\zerovec$) the one(s) that minimize the average squared off-diagonal elements of $\Smat=\X^T\X=(s_{ij})$. The unconditional $E(s^2)$, denoted $UE(s^2)$, criterion \citep{jones2014optimal} also minimizes the squared off-diagonals but includes the intercept column and drops the column balance constraint. 
Many other supersaturated design criteria have been put forth, including Bayesian D-optimal supersaturated designs \citep{Jones08} and model-robust supersaturated designs \citep{Jones2009}.

\cite{MarleyWoods10} and \cite{weese_etal2015} used the simulation-based approach and GDS to compare different SSD construction methods known at the time. They found no clear optimal design among the different SSDs with respect to power (i.e., true positive rate) and false discovery rates. Later, \cite{weese2017powerful, weese2021strategies} proposed and investigated the $Var(s+)$-criterion that minimizes the variance of the off-diagonal elements of $\Smat$ while ensuring a specified $UE(s^2)$ efficiency and forcing the average off-diagonal to be positive, $E(s)>0$. Their simulations under the GDS showed their SSDs increased power and decreased false discovery rate compared to existing designs when effect directions are known. This improvement was only noted under the GDS analysis, motivating \cite{singh2022selection} to identify SSDs lying on a Pareto front of two heuristic criteria derived from the theory of the Dantzig selector.  The designs of \cite{singh2022selection} are the first to connect the structure of $\X^T\X$ with a regularization method.  They showed via simulation that the PEDs were similar or better than the catalogued $Var(s+)$ designs in terms in power and false discovery rate.

Section \ref{sec:G-D_sims} provides more examples of simulation result differences due to tuning parameter selection when using regularization methods to analyze screening experiments. This sensitivity has reproducibility implications, especially if the description of the implementation of the GDS is not clearly stated. 




\subsection{GDS and Screening Designs with Interactions}

Two-level screening designs that entertain two-factor interaction effects have been studied extensively \citep[e.g.,][]{Mee,mee2017selecting}, including for example, Bayesian D-optimal designs \citep{dumouchel_jones1994, Jones08}, orthogonal arrays \citep{hedayat1999orthogonal}, and model robust screening designs \citep{li_nachtsheim2000, loeppky2007nonregular, smucker2012model, smucker2015approximate}. These designs allow for estimation of all main effects and typically provide for estimation of some two factor interactions. Often, empirically-based assumptions are made about $\betavec$ when constructing and analyzing the designs. The effect and factor sparsity principles say that relatively few of the $p=k+k(k-1)/2$ effects in the model are truly active, and that the active effects correspond to only a few of the $k$ factors. Effect hierarchy suggests only a small proportion of the interaction effects are active, and that their magnitude is typically smaller than that of the active main effects \citep{li2006regularities,ockuly2017response}. Given these assumptions, it is common practice to compare designs by partitioning estimation of $\mathcal{A}$ into estimation of the active main effects and estimation of active interactions.


\cite{mee2017selecting} recently compared two-level screening designs using the GDS, and \citet{draguljic_etal2014} used several regularization methods including the lasso and the GDS.  Others have used GDS to analyze three-level designs that can entertain quadratic effects. \cite{hameed2022analysis} used the GDS as a baseline analysis method in their analysis of orthogonal minimally aliased response surface designs.  \cite{weese2018analysis} and \cite{errore2017using} use the same approach to analyze definitive screening designs. \cite{vazquez2021mixed} use the lasso and the Dantzig selector as a baseline comparison for mixed level designs. Again, Table~\ref{tab:dant_param} shows the varying implementation of GDS for these papers' simulation studies.

\subsection{Paper Overview}
The paper is organized as follows. Section~\ref{sec:G-D_sims} illustrates the different conclusions reached for SSDs for different implementations of the GDS through simulation. It considers sensitivity for both SSDs and screening designs with interactions. Section~\ref{sec:exact_prob} introduces a new graphical analysis of penalized regression models based on simulated and exact screening probabilities, with particular attention give to the lasso. In  Section~\ref{sec:comparing_exact}, we use the proposed approach to compare designs, for both supersaturated and larger screening experiments. Lastly, we provide a discussion of our work in Section \ref{sec:discussion}.


\section{Simulations to Study Inconsistencies in the Gauss-Dantzig Selector} \label{sec:G-D_sims}

Figure~\ref{fig:coarse_fine} illustrated that differences can arise when the search grid for the Dantzig selector tuning parameter is varied. Since the implementation of the GDS also requires the choice of a model selection statistic and threshold parameter $\gamma$, it is worth investigating differences that arise due to those choices as well. Since the GDS has been utilized for both SSDs and larger screening designs that include two-factor interactions, in this section we make comparisons exclusively using the GDS. 

\subsection{GDS Inconsistencies in SSD Simulation Studies}\label{sec:marley_woods_sim}
To investigate the sensitivity of the GDS in the context of SSDs, we repeat the simulation studies presented in the seminal work of \cite{MarleyWoods10} and vary the choice of the selection statistic, threshold parameter ($\gamma$), and the search grid for $\lambda$. This allows us to observe how conclusions reached in \cite{MarleyWoods10} would change with the simulation settings. The original goal of the \cite{MarleyWoods10} study was to compare SSD construction and analysis methods to provide design size implications regarding sparsity assumptions to SSD users.

\cite{MarleyWoods10} use three different SSD sizes ($n$, $p$): (12, 26), (14, 24) and (18, 22) and two types of SSDs, Bayesian D-optimal \citep{Jones08} and balanced $E(s)$-optimal. We consider the same sparsity and effect size scenarios as \cite{MarleyWoods10}, and set $\gamma$ at either $\gamma=\sigma$  or a data driven threshold of $\gamma=0.1*\text{max}|\hat{\betavec}|$.  We also consider both AICc and BIC as model selection statistics and assume effect directions are unknown.  Lastly we experiment with a fine and coarse search grid for $\lambda$ as described in Section~\ref{sec:Introduction}.  The true models were generated according to \eqref{eq:me_model} with $\sigma^2=1$ and the truly inactive factors are assigned $\beta_j=0$.

\begin{figure}[ht]
    \centering
    \includegraphics[width=5in]{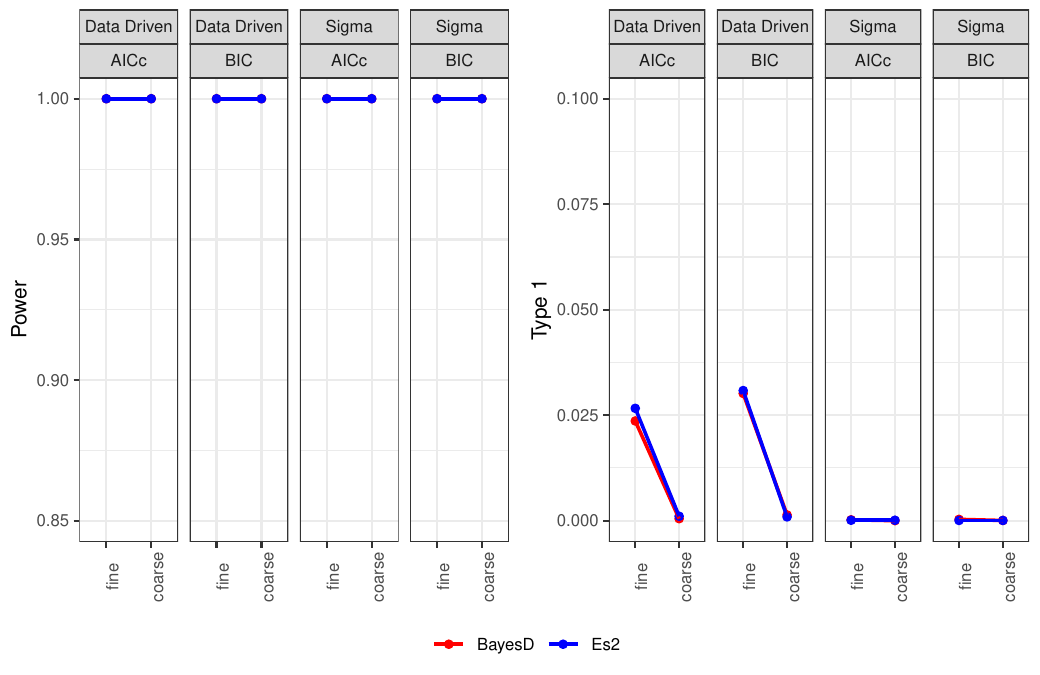}
    \caption{Comparison of (18, 22) Bayesian D-optimal and balanced $E(s^2)$-optimal SSDs for $a=3$ active effects with a magnitude of 5 and unknown true effect signs.  Note the difference in the y-axis values in the two panels.}
    \label{fig:n18k22_c3mu5}
\end{figure}

For high sparsity, large effect scenarios, the choice of the tuning parameters generally made little to no difference in the conclusions regarding SSD choice.  For example, Figure~\ref{fig:n18k22_c3mu5} shows the results for the (18, 22) SSD in the most sparse, largest effect scenario considered by \citet{MarleyWoods10} (3 active factors of magnitude $\pm5\sigma$). Clearly, in this case, any perceived differences in designs are inconsequential. Regardless, it is interesting to note the clear impact of search grid choice on Type I error values when making use of a data driven threshold.   

\begin{figure}[ht]
    \centering
    \includegraphics[width=5in]{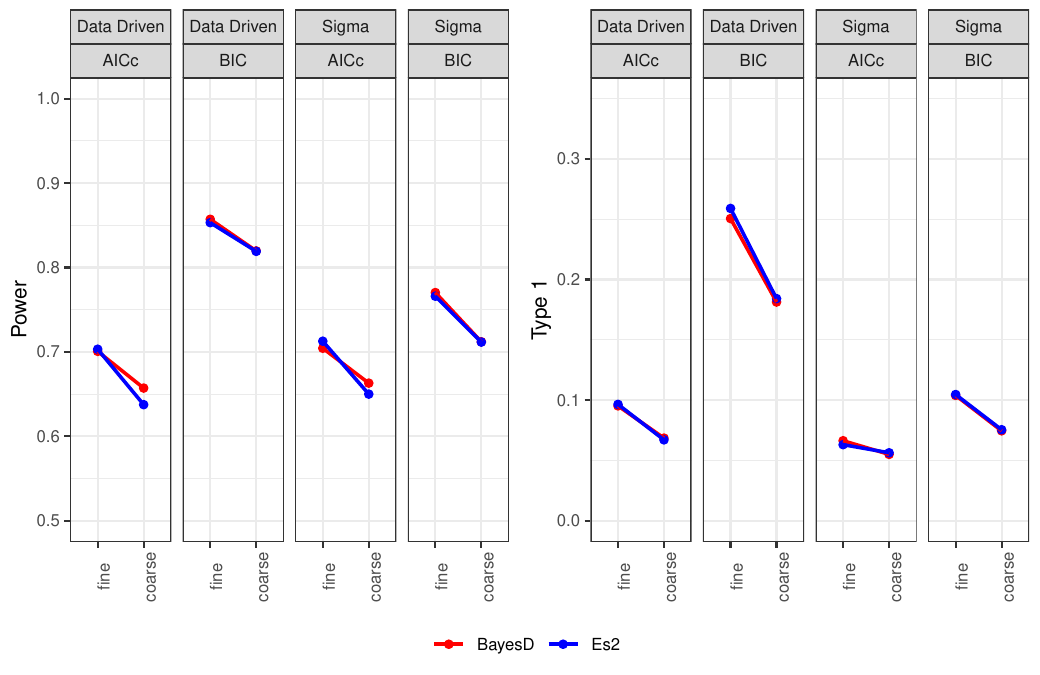}
    \caption{Comparison of (14, 24) Bayesian D-optimal and balanced $E(s^2)$-optimal SSD for $a=6$ active effects with a magnitude of 3 and unknown true effect signs.}
   \label{fig:n14k24_mixed}
\end{figure}

Figure~\ref{fig:n14k24_mixed} considers a more challenging scenario with the case of a (14, 24) SSD and six active factors (each having a magnitude of $\pm3\sigma$). In this case, the granularity of the tuning grid will change the conclusion regarding the preferred SSD with respect to power if AICc is used as the model selection statistic. Furthermore,  differences in power and Type I error values are evident when using a fine vs. coarse search grid as well as AICc vs. BIC and/or a data driven choice of $\gamma$ vs. $\gamma=\sigma$. These differences alone highlight the need for consistency in published simulation studies.    

\begin{figure}[ht]
    \centering
    \includegraphics[width=5in]{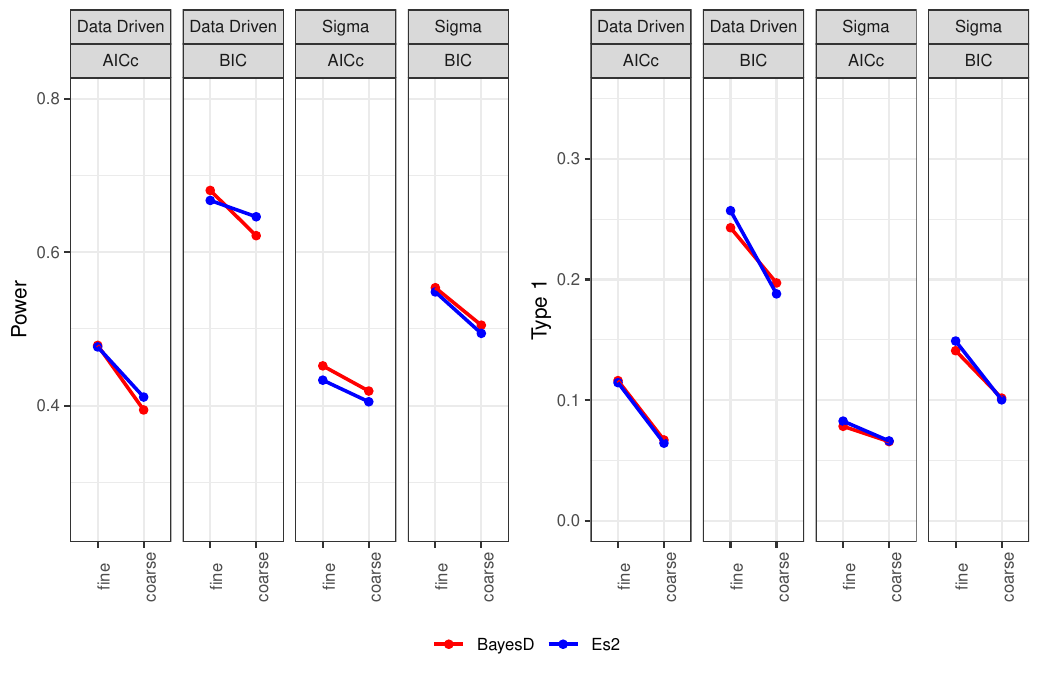}
    \caption{Comparison of (12, 26) Bayesian D-optimal and balanced $E(s^2)$-optimal SSD for $a=6$ active effects with a magnitude of 3 and unknown true effect signs.}
    \label{fig:n12k26_mixed}
\end{figure}

Finally, Figure~\ref{fig:n12k26_mixed} shows results of the (12, 26) case (six active factors each having magnitude $\pm3\sigma$). In this case, there is potential confusion regarding design preference when making use of the data driven threshold and BIC. With the coarse grid one would conclude that the $E(s^2)$-optimal design was preferred, with both higher Power and lower Type 1 error rate, whereas the script is flipped for the fine grid.

Although not shown in the body of paper, the results of the complete set of simulations based on \cite{MarleyWoods10} are in the supplementary materials. 
It is conceivable that \cite{MarleyWoods10} might have reached alternate conclusions if different tuning parameter selection strategies were used.  

\subsection{GDS Inconsistencies with Larger Screening Designs}\label{sec:larger_designs}



\cite{mee2017selecting} compare two-level screening designs based on power and false discovery rates.  They use a variety of screening designs (i.e. orthogonal arrays, Bayesian $D$-optimal, model robust), analyzed via the Dantzig selector and forward selection.  The authors implement the GDS using the automated approach as described in Section~\ref{sec:Introduction} with the AICc statistic, a threshold value of $\gamma=0.5$ and a search grid that steps from 0 to $\max(\X'\yvec)$ by 1.  To quantify the impact of these choices on the outcome of the design comparison study, we repeat the simulation scenarios described in \cite{mee2017selecting} using the fine and course grids described in Section~\ref{sec:Introduction}, as well as both AICc and BIC. We keep $\gamma=0.5$ as in the original work.  As in \cite{mee2017selecting} we generate the response according to model~\eqref{eq:me_model}. Coefficients for the truly active main effects are sampled (with replacement) from $\{2,2.5,3,3.5\}$. Coefficients for truly active two-factor interactions are sampled from either $\{0.5,1,1.5,2\}$ or $\{2, 2.5, 3, 3.5\}$ for two-factor interaction coefficients that are either smaller or equivalent, respectively, to the truly active main effect coefficients.  All signs were assumed to be unknown, and thus randomly assigned to each parameter with equal probability.  \cite{mee2017selecting} compared the power and false discovery rate (FDR) for the subsets of main effects and two-factor interactions separately and we will follow suit.  We highlight only two particularly interesting parts of the results here; the entire simulation for the $n=20$ and $k=7$ designs are provided in the Supplementary Materials. 


Figure~\ref{fig:large2fi_grid} shows the case when the two-factor interactions are equal in magnitude to the main effects, measuring the differences in FDR between the different $\lambda$ search grids (``Fine'' and ``Coarse'') and different selection statistics (AICc and BIC). Note that each line in the plots represent a different design from Figure 3d in \citet{mee2017selecting}.  In this case the model selection statistic does not appear to drastically impact the FDR values. However, the choice of the search grid shows different FDR behavior across the number of active two-factor interactions. The FDR is seemingly constant for the finer search grid and steadily increasing for the coarse search grid. 

\begin{figure}[ht]
    \centering
    \includegraphics[width=4in]{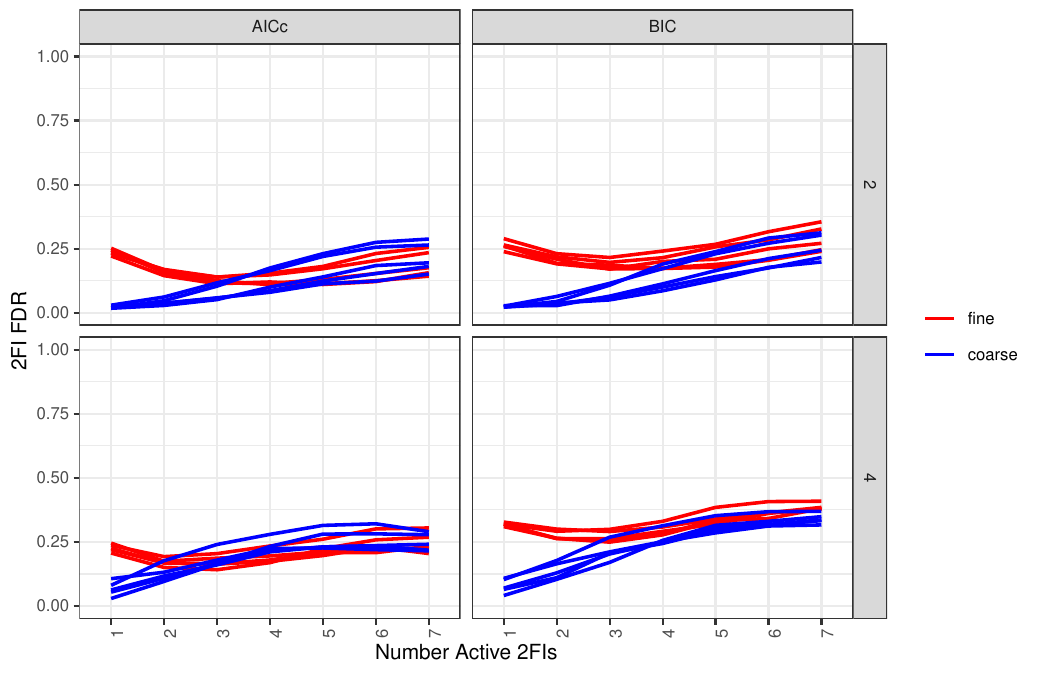}
    \caption{Comparison of $n=20$, $k=7$ designs from Figure 3d in \citet{mee2017selecting} using the Dantzig selector, comparing AICc and BIC and the previously defined ``Fine'' and ``Coarse'' search grids, with $\gamma=0.5$.  The FDR is compared for 2 and 4 active main effects and 1 to 7 active two-factor interactions.}
    \label{fig:large2fi_grid}
\end{figure}

\begin{figure}[ht]
    \centering
   \includegraphics[width=5in]{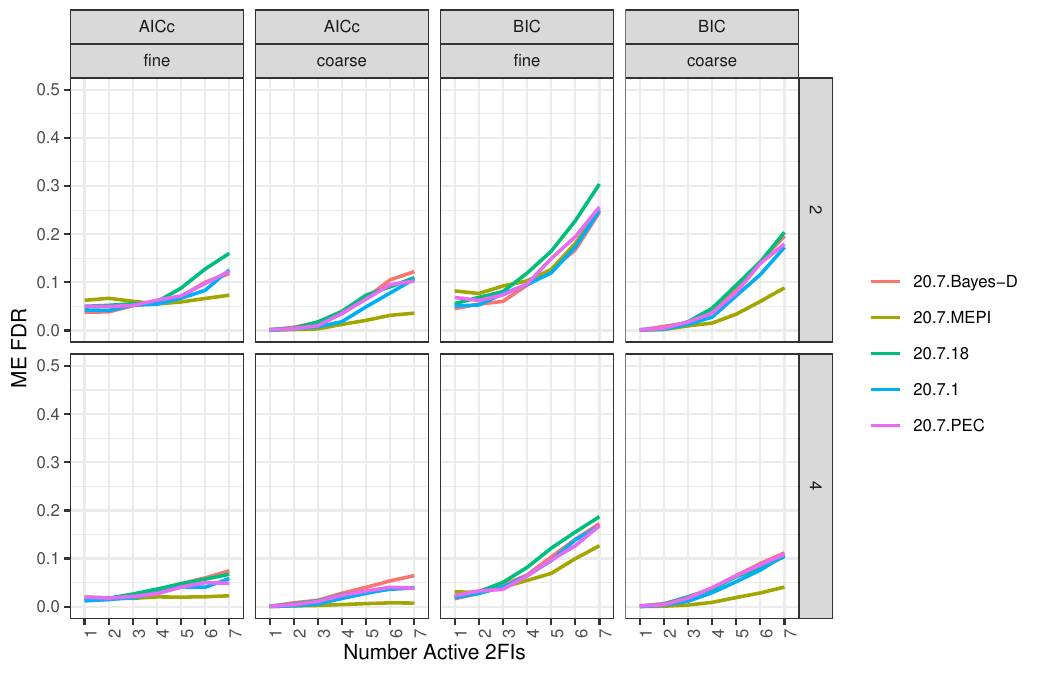}
   \caption{Comparison of $n=20$, $k=7$ designs from Figure 3d in \citet{mee2017selecting} using the Dantzig selector, and comparing AICc and BIC as well as a ``Fine'' and ``Coarse'' search grid with $\gamma=0.5$.  The FDR is compared for 2 and 4 active main effects and 1 to 7 active two-factor interactions.}
   \label{fig:meFDR}
\end{figure}

Figure~\ref{fig:meFDR} illustrates a case where design preference conclusions might change based on the choice of selection statistic.  \cite{mee2017selecting} conclude that for the $n=20$ and $k=7$ designs the MEPI design has good main effect power and the lowest FDR for main effects.  Figure~\ref{fig:meFDR} shows that if the authors had chosen to use a finer grid with the BIC statistic, the MEPI design does not show a better FDR for two active main effects and is slightly better when there there are four active main effects. 

\section{A Graphical Comparison of Screening Probabilities}\label{sec:exact_prob}

As noted, simulation studies to compare designs under a regularization method currently require specifying several aspects related to the penalty parameter, $\lambda$, along with the specification of a threshold $\gamma$. 
We have demonstrated that the comparisons can be sensitive to different choices of these aspects. What is perhaps even more concerning is that the simulation analysis is inconsistent with the recommended analysis for an actual experiment: graphically examine the solution path across the $\lambda$ regularization parameter. Thus, we propose a graphical method for estimating $\mathcal{A}$ in screening experiments which summarizes a design's solution path in terms of its support recovery probabilities, $P(\hat{\mathcal{A}} = \mathcal{A})$, where $\hat{\mathcal{A}}$ is the support from $\hat{\betavec}_{DS}$ (or any regularized estimator). 

\subsection{Simulation-Based Support Recovery Probabilities}

To compare designs, we recommend estimating these probabilities across a dense grid of $\lambda$ via simulation similar to Section~\ref{sec:G-D_sims}, and then plotting these probabilities. 
If these probabilities are high for a wide range of $\lambda$, then the design should consistently produce solution paths that accurately estimate $\mathcal{A}$.


To begin, we fix $a$ (the number of active effects), set $\sigma^2=1$, and propose a distribution for the magnitude and sign of the active effects. Given these settings, simulate $niter$ data sets, producing $niter$ solution paths for $\hat{\betavec}_{DS}$ (or any regularized estimator) across a fine grid of $\lambda$. The solution paths are then checked at each $\lambda$ value for whether or  not they achieve support recovery, producing a vector of 0's and 1's where 1 indicates $\hat{\mathcal{A}}=\mathcal{A}$. Averaging these $niter$ vectors of $0$'s and $1$'s provides a vector of estimated probabilities of support recovery. This approach is no more computationally demanding than the simulations in Section~\ref{sec:G-D_sims}. The important difference is that 
we retain more information about the screening behavior by not requiring specification of a thresholding parameter nor a tuning parameter selection method.

\begin{figure}[ht]
    \centering
    \includegraphics[width=5in]{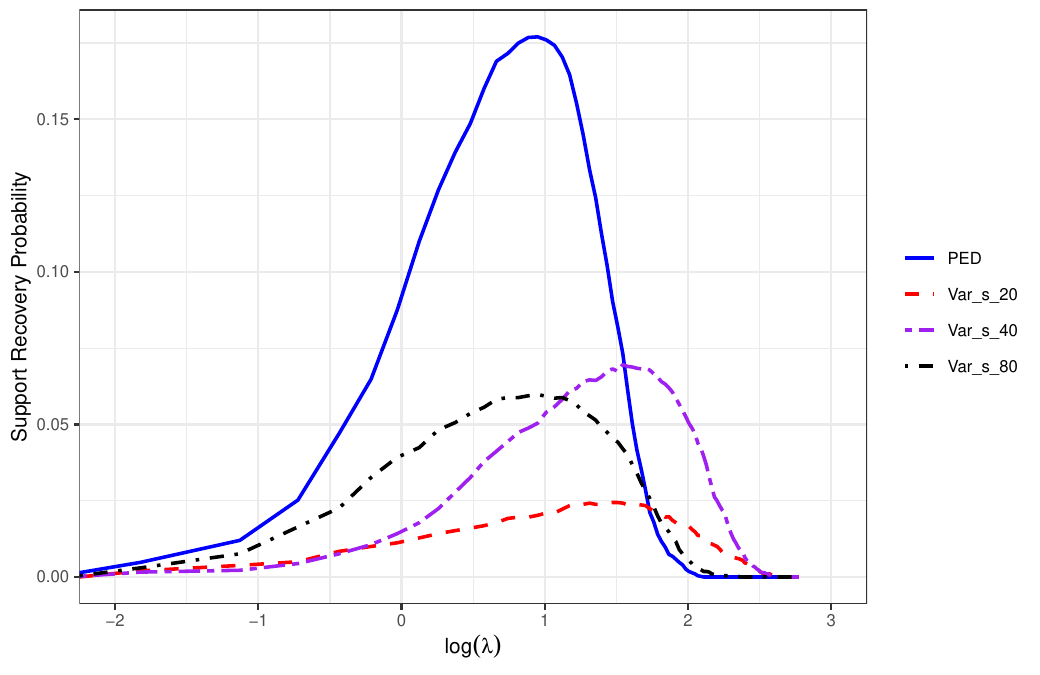}
   \caption{Comparison of $n=14$ and $k=24$ PED and three $Var(s+)$ designs using simulated Dantzig seletor support recovery probability. We assumed $a=7$ active effects randomly drawn from $Exp(1)+3$ for each iteration. The three $Var(s+)$ designs correspond to three $UE(s^2)$-efficiency constraints: 80\%, 40\%, and 20\%.}
   
    \label{fig:phi_comp_example}
\end{figure}

To demonstrate this approach, we revisited
the simulation study from Figure~\ref{fig:coarse_fine} with $SN=3$ and $n/2=7$ active factors. Under a coarse tuning parameter grid, the PED had higher power and nearly identical Type 1 error compared to the $Var(s+)$ design; the fine tuning parameter grid implied these designs were on more equal footing with respect to power and Type 1 error. Figure \ref{fig:phi_comp_example} shows our graphical comparison approach for these two designs as well as two other $Var(s+)$ designs that reduce the $UE(s^2)$ efficiency constraint from 80\% to 40\% and 20\% (doing so should increase the pairwise column correlations). All designs exhibited low overall support recovery probabilities, implying that most solution paths generated from these designs will not perfectly identify the support. However, this is not surprising, since Figure~\ref{fig:coarse_fine} showed that the high power was accompanied by a Type 1 error rate around $0.20$. When it comes to relative design comparisons, the PED shows clear improvement in support recovery probability compared to the $Var(s+)$ designs across nearly all $\lambda$ values. The maximum support recovery probability for the PED is estimated to be $0.175$, occurring at $\log(\lambda)\approx 1$.  Ranking the 80\% and 40\% efficient $Var(s+)$-designs is not as straightforward. The 40\% efficient design has a higher maximum probability, but the $80$\% efficient design has a maximum probability nearly as high and sustains a higher probability for smaller $\lambda$ values.
Smaller $\lambda$ values introduce less penalization in the estimates, and hence reduce bias. Therefore, it is desirable to have larger support recovery probabilities at smaller $\lambda$ values.  In general, we should expect the solution paths for the $80$\% efficient $Var(s+)$ designs to better capture the true support than the 40\% efficient design.

\subsection{Exact Sign Recovery Probabilities for the Lasso}

Although straightforward, one drawback of the above simulation-based approach is that $niter$ needs to be large for the estimated probabilities to be accurate and reproducible. Direct calculations, when available, are preferred. Recently, \citet{stallrich2023optimal} developed an optimal design framework that targets the probability of support recovery for the lasso. Lasso estimates are the solution to 
\begin{align}\label{eq:lasso}
   \hat{\betavec} = \arg\min_{\betavec}  \frac{1}{2n}(\yvec - \X\betavec)^T(\yvec - \X\betavec) + \lambda \|\betavec\|_1.
\end{align}
The lasso has been used to analyze SSDs \citep{draguljic_etal2014, zhang2007method, koukouvinos2008method} but it is not as common as the Dantzig selector in the SSD literature, even though its performance is quite similar \citep{draguljic_etal2014}. The lasso is well suited for problems where $n < p$ and its properties are well known \citep{tibshirani2012lasso, zhao2006model, jia2015preconditioning}. There are strong theoretical \citep{meinshausen2007discussion, lounici2008sup, bickel2009simultaneous, james2009dasso, asif2010lasso} and empirical \citep{draguljic_etal2014} links between the Dantzig selector and lasso, so this approach is a reasonable proxy for designs analyzed by either method. We adopt this exact probability approach for the remainder of the paper.

Technically, the criteria in \cite{stallrich2023optimal} look to maximize the probability the lasso estimator recovers the sign of $\betavec$, which has a simple, closed-form expression \citep[for more details, see the Supplementary Materials and][]{stallrich2023optimal}. Sign recovery implies support recovery, requiring that all inactive effects are estimated as zero and all active effects are estimated with the correct sign. \cite{stallrich2023optimal} consider these probabilities for the cases of known and unknown effect directions.  For known signs, one can assume without loss of generality that the active effects are all positive. For unknown signs, they assume each of the $2^a$ possible sign vectors are equally likely. The probability of sign recovery for a known sign vector with $a$ active effects and a given $\betavec$ is denoted by $\Phi_{\lambda}(\X \, | \, a, \betavec)$. This probability considers all possible supports of size $a$ and all permutations of the active effects of $\betavec$. In the case of unknown signs, they denote the probability of sign recovery with $a$ active effects and a given $\betavec$ by $\Phi_{\lambda}^{\pm}(\X \, | \, a, \betavec)$. 

In their optimal design framework, \cite{stallrich2023optimal} use a user-defined minimum signal to noise ratio for the active effects, denoted $\beta$, and set the absolute magnitude of all active effects to $\beta$. If a distribution of active effects were desired instead (as in Figure~\ref{fig:coarse_fine}), one would need to average the sign recovery probabilities across this  $\betavec$ distribution. To estimate this averaged probability, we recommend drawing $niter$ $\betavec$'s, say $\betavec_1,\dots,\betavec_{niter}$, and then average these probabilities across the grid of $\lambda$ values. We denote the two sets of probabilities by 
\begin{equation}
    \Phi_\lambda( \X \, | \, a)=\frac{1}{niter}\sum_i \Phi_{\lambda}(\X \, | \, a, \betavec_i) \qquad \Phi_\lambda^{\pm}( \X \, | \, a)=\frac{1}{niter}\sum_i \Phi_{\lambda}^{\pm}(\X \, | \, a, \betavec_i)\ .\
\end{equation}

The exact calculation of the lasso sign recovery measure is computationally feasible when comparing SSDs, the primary use-case considered by \cite{stallrich2023optimal}. 
Calculating the exact lasso sign recovery probability becomes much more cumbersome in the presence of two-factor interactions due to the number of possible submodels. In this scenario, we revert back to the original simulation-based approach that began this section to estimating $\Phi_{\lambda}$, replacing the Dantzig selector with the lasso, and support recovery with sign recovery. In our implementation, we randomly select a set of submodels with $k$ main effects and $g$ two-factor interactions (under, say, a weak-heredity assumption). Next, the effect sizes for each active main effect and two-factor interaction are randomly generated under some distribution. The exact sign recovery probabilities for each of the $niter$ randomly generated supports can then be calculated across the grid of $\lambda$ and then averaged to estimate $\Phi_{\lambda}$ or $\Phi_{\lambda}^{\pm}$. 
Simulation of the lasso sign recovery probability in this way allows for the evaluation of the lasso sign recovery among the main effects and two-factor interactions separately, which is consistent with how larger screening designs have been evaluated with other penalized estimation approaches.


\section{Examples} \label{sec:comparing_exact}

We now revisit design comparisons from previous sections and compare designs using the graphical $\Phi_\lambda$ method from Section~\ref{sec:exact_prob}. 
We start with comparing SSDs using the exact $\Phi_\lambda$ probability calculations. Then we illustrate the simulated lasso sign recovery method for larger screening designs that consider detection of two-factor interactions.

\subsection{Comparing SSDs Using Exact Screening Probabilities}\label{sec:compare_ssd}
 
\begin{figure}[ht]
    \centering
    \includegraphics[width=6in]{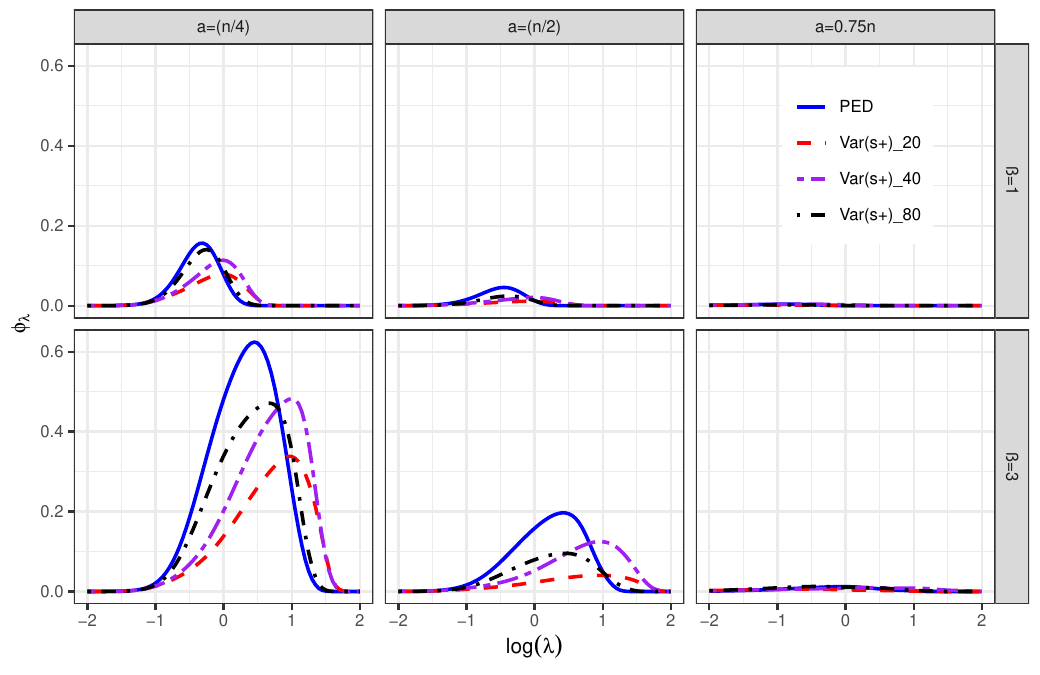}
   \caption{Graphical comparison of the PED and $Var(s+)$-optimal designs referenced in Figure \ref{fig:coarse_fine}. Note here the active effects were fixed to be either $\beta=1$ or $3$. }
   
    \label{fig:PED_vars_phi}
\end{figure}

Figure~\ref{fig:phi_comp_example} revisited one scenario from Figure~\ref{fig:coarse_fine}, demonstrating a more obvious improvement of the PED over the $Var(s+)$ design. Using exact lasso screening probabilities, Figure~\ref{fig:PED_vars_phi} shows the $\Phi_\lambda$ performance for the four considered designs from Figure~\ref{fig:phi_comp_example} over a range of $\log(\lambda)$ for a known sign vector for two fixed values $\beta=1,\, 3$ and three support sizes $a=4,\, 7,\, 11$. Across all situations, the PED demonstrates higher lasso sign recovery performance over a larger $\log(\lambda)$ range compared to the $Var(s+)$-optimal design. Interestingly, while \citet{weese2017powerful} stated that the efficiency constraint has little impact on screening effectiveness, except at the extremes, Figure \ref{fig:PED_vars_phi} demonstrates clear differences in lasso sign recovery performance as we vary the $UE(s^2)$ efficiency. However, the 80\% efficient $Var(s+)$ design consistently has higher sign recovery probabilities than the other $Var(s+)$ designs for smaller $\lambda$.


The bottom middle panel of Figure~\ref{fig:PED_vars_phi} most closely resembles the settings from Figure~\ref{fig:phi_comp_example}, with the latter visualizing support recovery probability of the Dantzig selector. The behavior of the two plots is nearly identical (note the difference in scale between the two plots), lending support that one could anticipate behavior of the Dantzig selector's support recovery probability using the sign recovery probability of the lasso. Finally, we must also point out the overall poor performance of the designs when $a=0.75n=11$. The graphical analysis implies that it is highly unlikely for the solution path analysis to reveal the true support. Therefore, if a practitioner anticipates a large number of active effects, they should consider either reducing the number of factors considered, or increase their run size.

\subsection{Comparing Larger Screening Designs via Sign Probabilities}\label{sec:sign_prob_sim}

\begin{figure}[ht]
    \centering
   \includegraphics[width=6in]{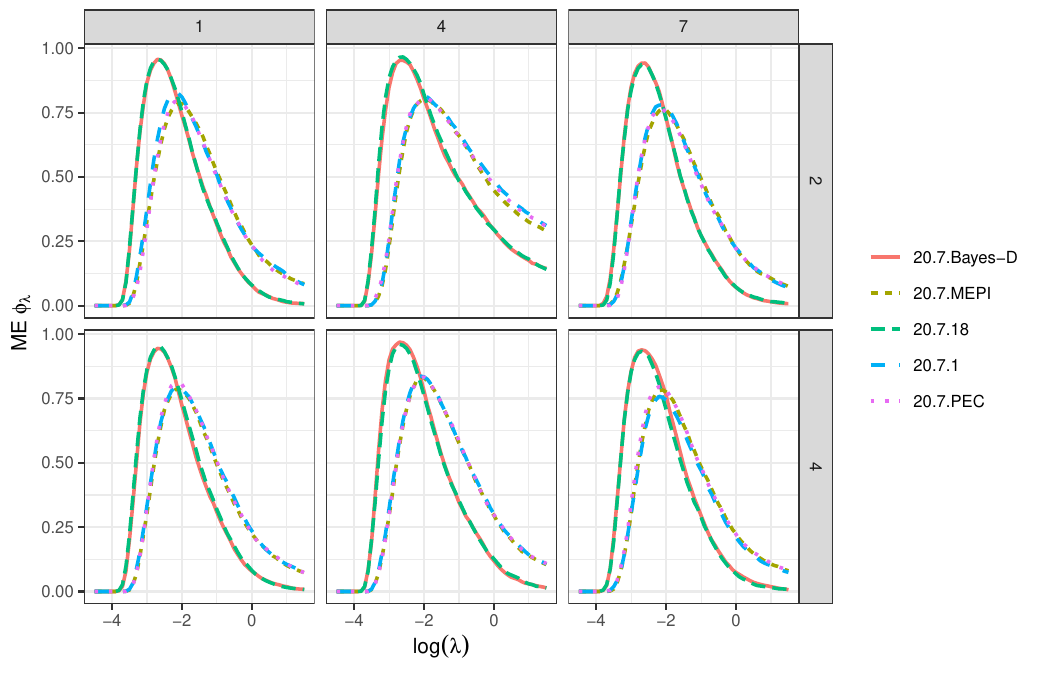}
   \caption{Comparison of $n=20$, $k=7$ designs using the simulated lasso sign recovery probability for the main effects only.  The sign recovery probability for main effects is compared for 2 and 4 active main effects (rows in the figure) and 1, 4, and 7 active two-factor interactions (columns in the figure).}
   \label{fig:lasso_sim_ME}
\end{figure}

In the case where $n>k+1$ and two-factor interactions are considered, the simulation based procedure described in Section \ref{sec:exact_prob} can be used to estimate the lasso sign recovery probability. We demonstrate this approach to compare the $n=20$ and $k=7$ two-level designs in \cite{mee2017selecting}. Here, we simulated the lasso sign recovery probabilities in a case where the truly active two-factor interactions are smaller in magnitude than the main effects. Main effect sizes were selected from the set $\{-3.5,3.5\}$ to help highlight design differences. Two-factor interactions were sampled with replacement from $\{0.5, 1, 1.5, 2\}$ with signs $+/-$ assigned randomly according to \cite{mee2017selecting}. 
The simulated lasso sign recovery values, separated for main effects and two-factor interactions, are compared for 2 and 4 active main effects (see rows of the Figure~\ref{fig:lasso_sim_ME}) and 1 to 7 active interactions assuming weak heredity. For brevity, the results for 1, 4, and 7 active two-factor interactions are displayed. The others are available in the Supplementary Materials.

Figure \ref{fig:lasso_sim_ME} shows the simulated lasso sign recovery probability across values of $\log(\lambda)$ for main effects only. For each scenario in Figure \ref{fig:lasso_sim_ME}, the Bayesian $D$-optimal and 20.7.18 orthogonal array designs show larger main effect sign recovery probabilities, specifically at lower values of $\log(\lambda)$ compared to the other designs compared. While the 20.7.MEPI, 20.7.1 OA, and 20.7.PEC designs do see higher main effect sign recovery probabilities for larger $\log(\lambda)$ values, they do not achieve the maximum sign recovery probability. Thus, based on this measure of main effect screening capability, we prefer the Bayesian $D$-optimal design and the 20.7.18 orthogonal array over the others. 

\begin{figure}[ht]
    \centering
   \includegraphics[width=6in]{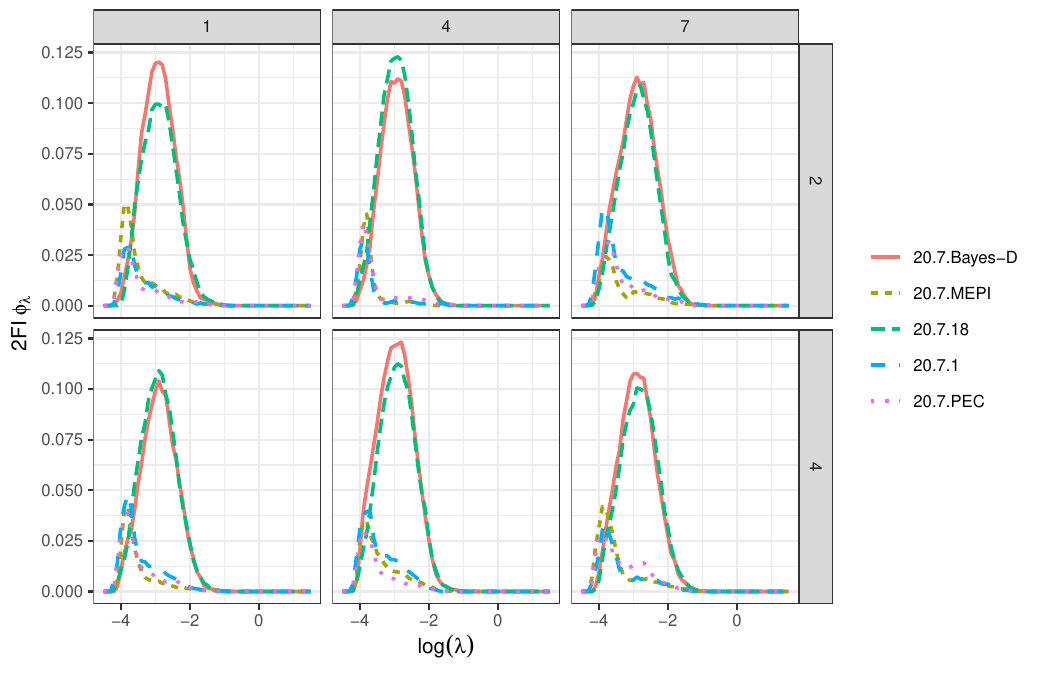}
   \caption{Comparison of $n=20$, $k=7$ designs using the simulated lasso sign recovery probability for the two factor interactions only.  The sign recovery probability for two factor interactions is compared for 2 and 4 active main effects (rows in the figure) and 1, 4, and 7 active two-factor interactions (columns in the figure).}
   \label{fig:lasso_sim_2FI}
\end{figure}

Figure \ref{fig:lasso_sim_2FI} shows the simulated lasso sign recovery probability across values of $\log(\lambda)$ for two-factor interactions only. Similar to  Figure \ref{fig:lasso_sim_ME}, the Bayesian $D$-optimal design and 20.7.18 orthogonal array show higher two-factor interaction sign recovery probabilities compared to the other designs for the scenarios in Figure \ref{fig:lasso_sim_2FI}. While the different scenarios show slight differences between the Bayesian $D$-optimal design and the 20.7.18 orthogonal array, neither stands out consistently as best.

The GDS simulation for the scenarios in \cite{mee2017selecting} did not show a dramatic difference between any of the five designs compared. However, the lasso sign recovery comparison shows a clear favoring of the Bayesian $D$-optimal design and the 20.7.18 orthogonal array over the MEPI, PEC, and 20.7.1 orthogonal array designs.   It should be noted that \cite{mee2017selecting} recommends the 20.7.1 orthogonal array design  over the others because of its overall performance when evaluated with the GDS and forward selection. 

\section{Discussion}\label{sec:discussion}
Regularization methods have become a popular tool to analyze screening designs, and to compare designs, especially in the SSD literature.  While the application of regularization methods is appropriate, we have illustrated potential reproducibility issues with tuning parameter selection. Including a limited number of possible tuning parameter values, as well as different model selection statistics can potentially lead to different conclusions regarding design choice and/or analysis. To combat this, we recommend comparing screening designs with exact lasso sign recovery probabilities for supersaturated designs, and estimating these probabilities for larger screening designs using simulation.

We have demonstrated a graphical method to compare these probabilities over a range of tuning parameter values, which circumvents the need to specify a grid to search over as in traditional simulation using regularization methods. Using sign recovery probabilities also requires no model selection statistic and no threshold choice. The graphical method of comparison allows for a much broader comparison of designs than that based on one or two metrics (e.g., power and type 1 error) conditioned on a set of tuning parameter choices. 
We have provided software for researchers and practitioners to use the lasso sign recovery probabilities to compare screening designs. For SSDs, the software computes $\Phi_\lambda$ and $\Phi_\lambda^{\pm}$ for a user input design, number of active effects, $\betavec$, and range of $\log(\lambda)$. For computationally burdensome scenarios, the package also has functionality to sample from the space of all possible active sets using the nearly balanced incomplete block design strategy in \cite{smucker2015approximate}. R software to calculate or simulate lasso sign recovery probabilities can be obtained here: https://github.com/kyoungstats/LassoSR.




We focused on plots of the lasso sign recovery probabilities because these probabilities have a closed form expression for known $\betavec$. As suggested in Section~\ref{sec:exact_prob}, a simulation-based approach could be employed to generate similar plots for other regularization methods (e.g., SCAD) and for other metrics of statistical quality, such as the probability the estimated support contains the true support, i.e., $P(\hat{\mathcal{A}} \subset \mathcal{A})$, or the mean-squared error loss, i.e., $E[(\hat{\betavec}-\betavec)^2]$. This is an area of future work.



\bibliographystyle{asa} 
\bibliography{power.bib}

\newpage
\begin{center}
{\large\bf SUPPLEMENTARY MATERIAL}
\end{center}

\setcounter{equation}{0}

\footnotesize

\section{Inconsistencies in SSDs}

This section provides the remainder of the simulations scenarios based on \cite{MarleyWoods10} from Section 2.1 in the main paper.  

\begin{figure}[H]
    \centering
    \includegraphics{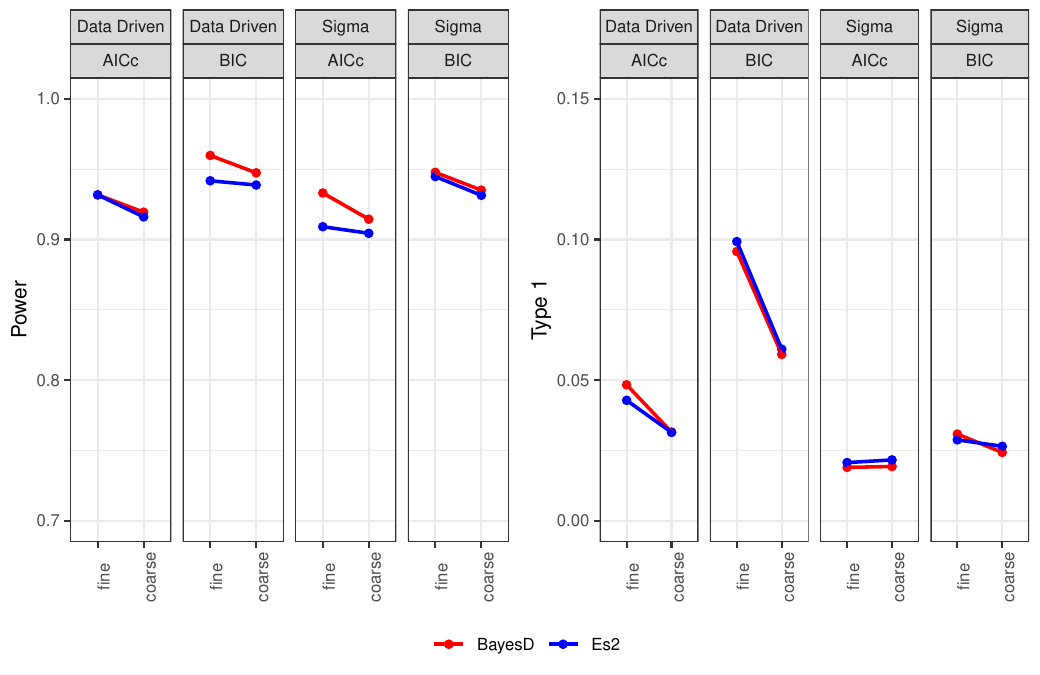}
   \caption{Comparison of (12, 26) Bayesian D-optimal and balanced $E(s^2)$ SSD for 3 active effects with an effect magnitude of 5 for unknown signs.}
   
    \label{fig:enter-label}
\end{figure}

 \begin{figure}[H]
     \centering
     \includegraphics{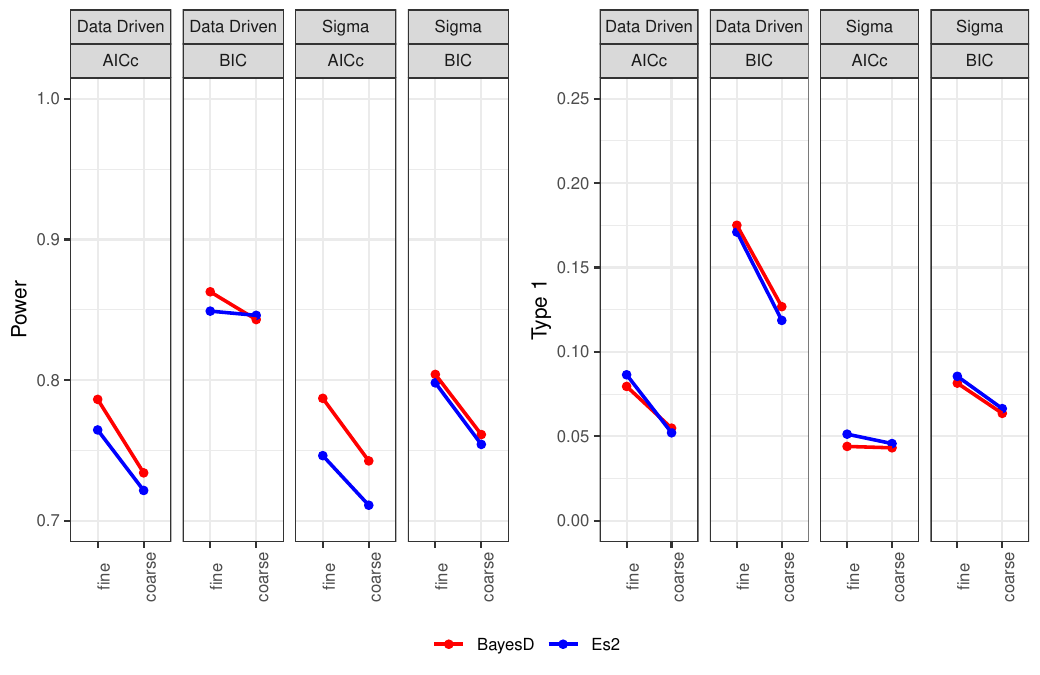}
     \caption{Comparison of (12, 26) Bayesian D-optimal and balanced $E(s^2)$ SSD for 4 active effects with an effect magnitude of 4 for unknown signs.}
     \label{fig:enter-label}
 \end{figure}

 \begin{figure}[H]
     \centering
     \includegraphics{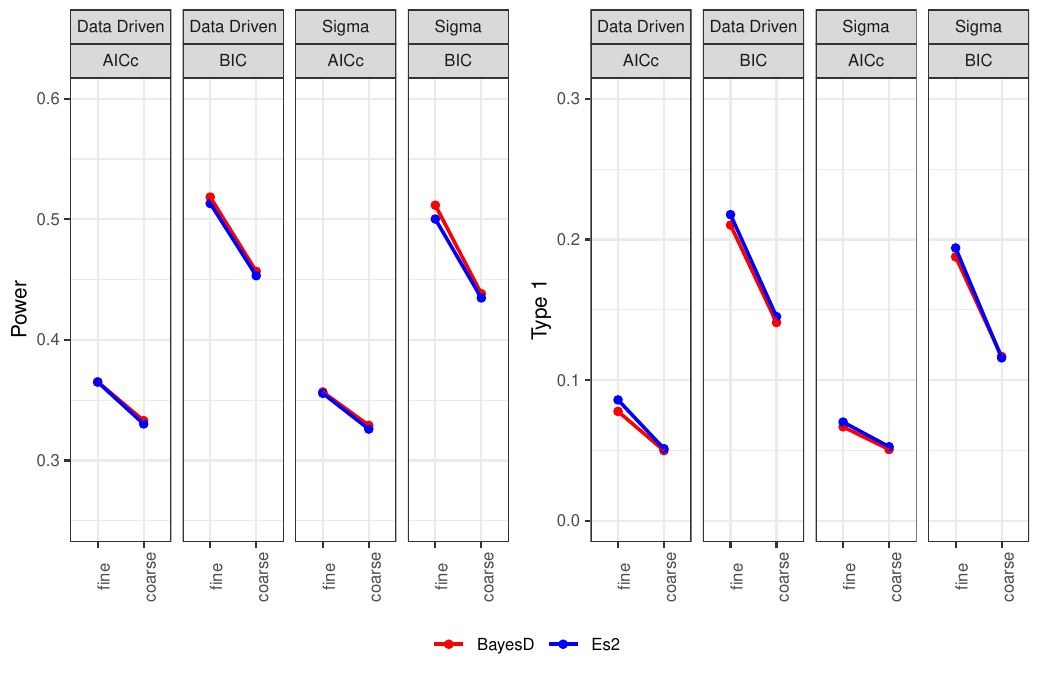}
     \caption{Comparison of (12, 26) Bayesian D-optimal and balanced $E(s^2)$ SSD for 9 active effects with an effect magnitude of the following values: 10,8,5,3 and the rest 2 for unknown signs.}
     \label{fig:enter-label}
 \end{figure}

 \begin{figure}[H]
     \centering
     \includegraphics{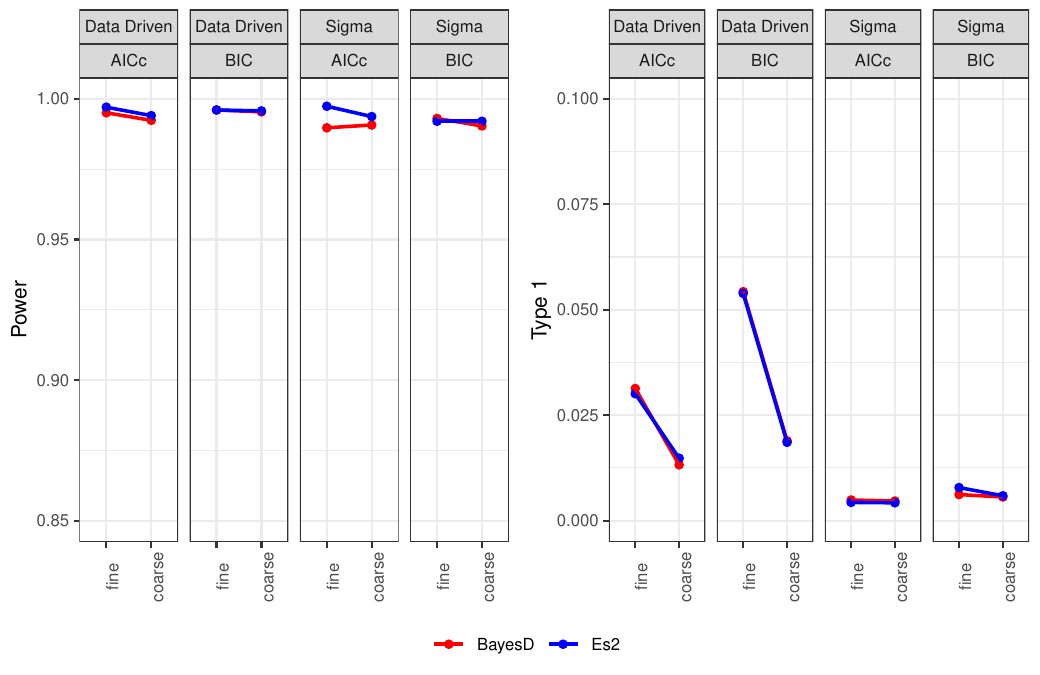}
     \caption{Comparison of (14, 24) Bayesian D-optimal and balanced $E(s^2)$ SSD for 3 active effects with an effect magnitude of 5 for both known and unknown signs.}
     \label{fig:enter-label}
 \end{figure}

 \begin{figure}[H]
     \centering
     \includegraphics{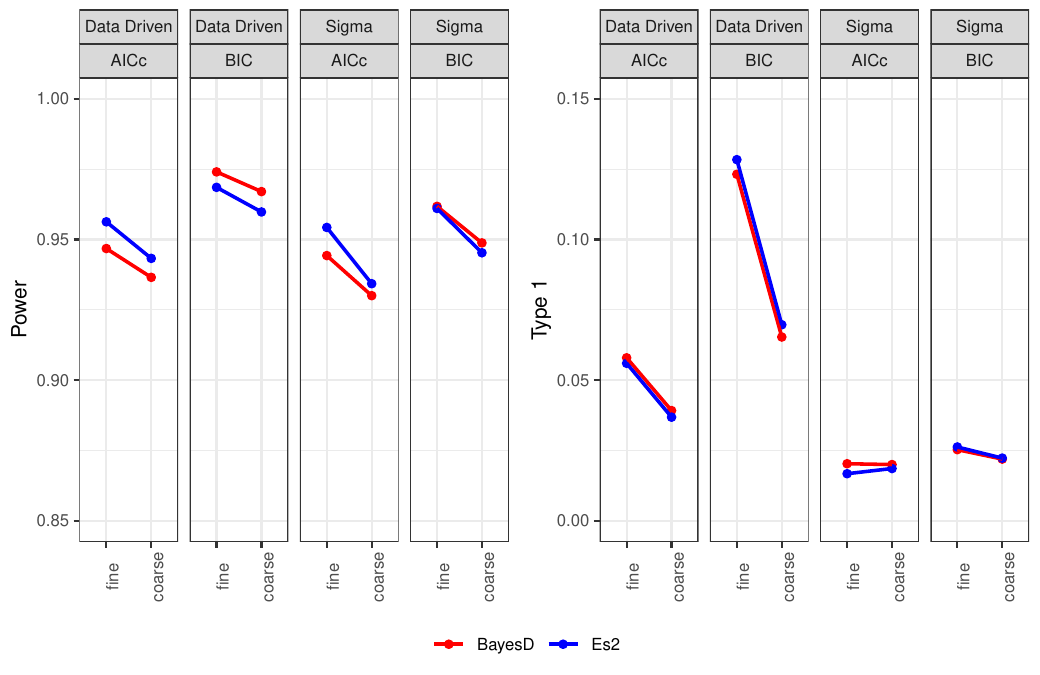}
    \caption{Comparison of (14, 24) Bayesian D-optimal and balanced $E(s^2)$ SSD for 4 active effects with an effect magnitude of 4 for both known and unknown signs.}
    \label{fig:enter-label}
 \end{figure}

 \begin{figure}[H]
     \centering
     \includegraphics{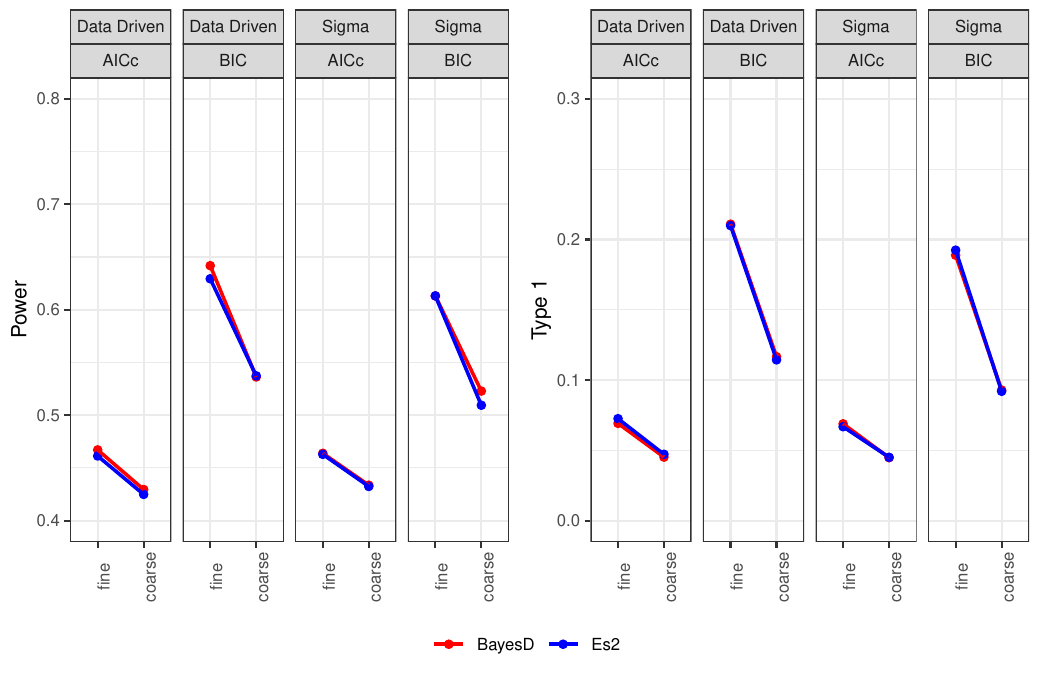}
     \caption{Comparison of (14, 24) Bayesian D-optimal and balanced $E(s^2)$ SSD for 9 active effects with an effect magnitude of the following values: 10,8,5,3 and the rest 2 for unknown signs.}
     \label{fig:enter-label}
 \end{figure}

 \begin{figure}[H]
     \centering
     \includegraphics{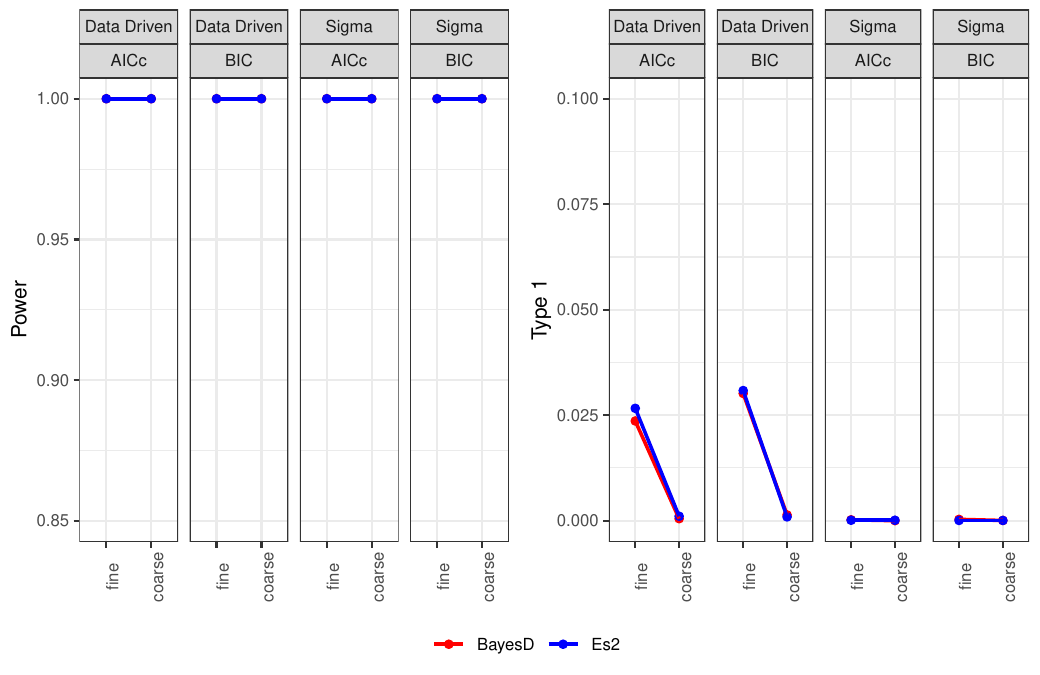}
     \caption{Comparison of (18, 22) Bayesian D-optimal and balanced $E(s^2)$ SSD for 3 active effects with an effect magnitude of 5 for unknown signs.}
     \label{fig:enter-label}
 \end{figure}

 \begin{figure}[H]
     \centering
     \includegraphics{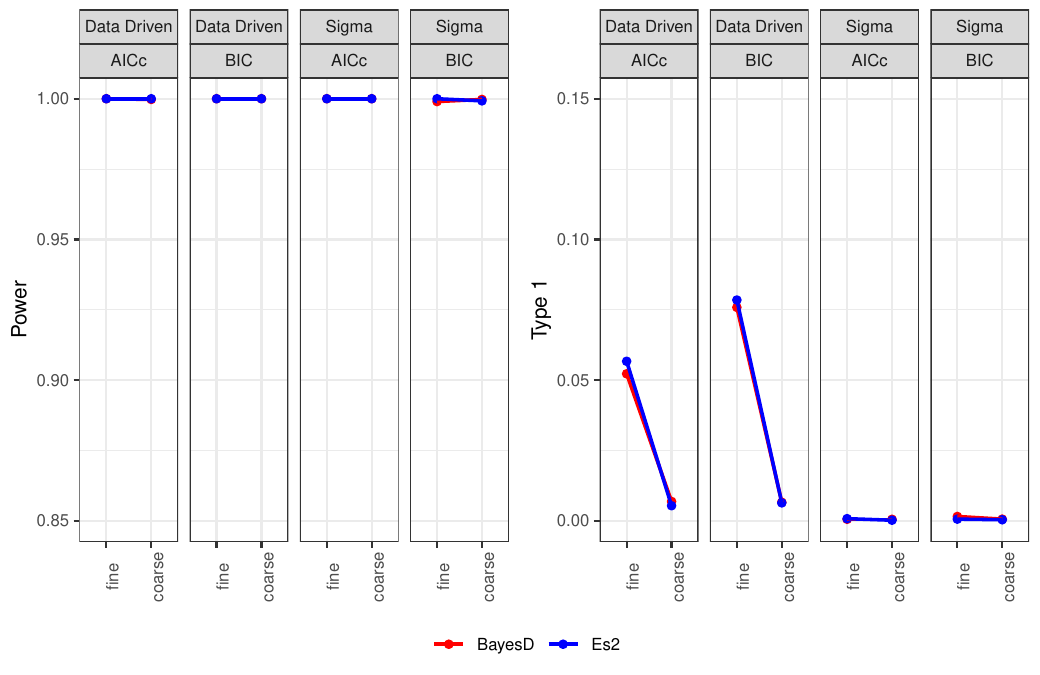}
     \caption{Comparison of (18, 22)  Bayesian D-optimal and balanced $E(s^2)$ SSD for 4 active effects with an effect magnitude of 4 for both unknown signs.}
     \label{fig:enter-label}
 \end{figure}

 \begin{figure}[H]
     \centering
     \includegraphics{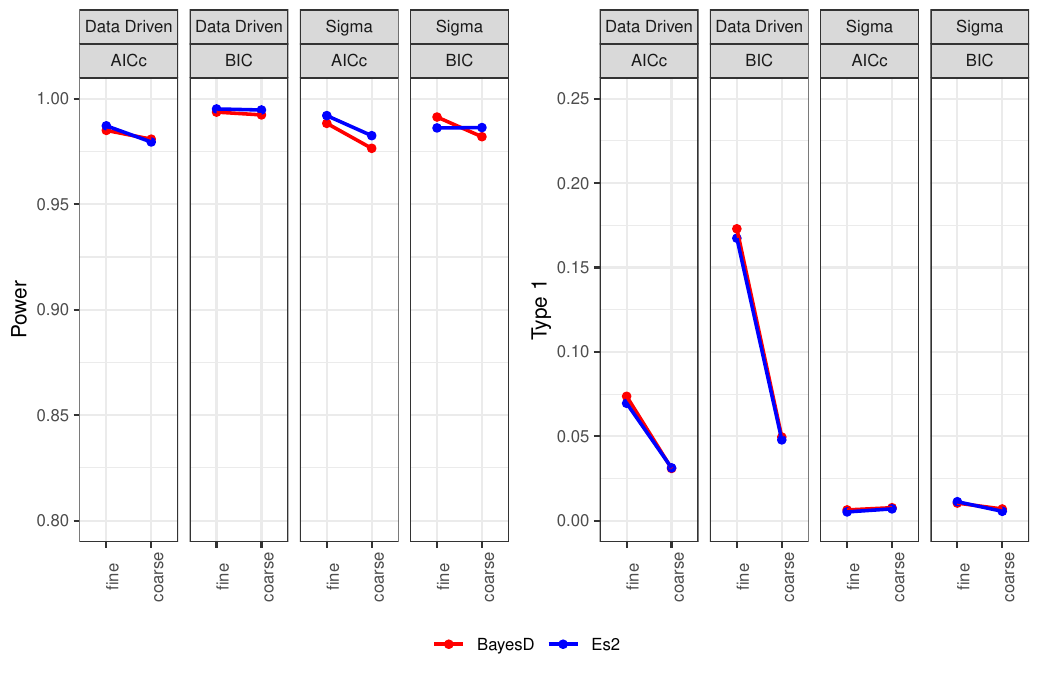}
     \caption{Comparison of (18, 22)  Bayesian D-optimal and balanced $E(s^2)$ SSD for 6 active effects with an effect magnitude of 3 for both unknown signs.}
     \label{fig:enter-label}
 \end{figure}

 \begin{figure}[H]
     \centering
     \includegraphics{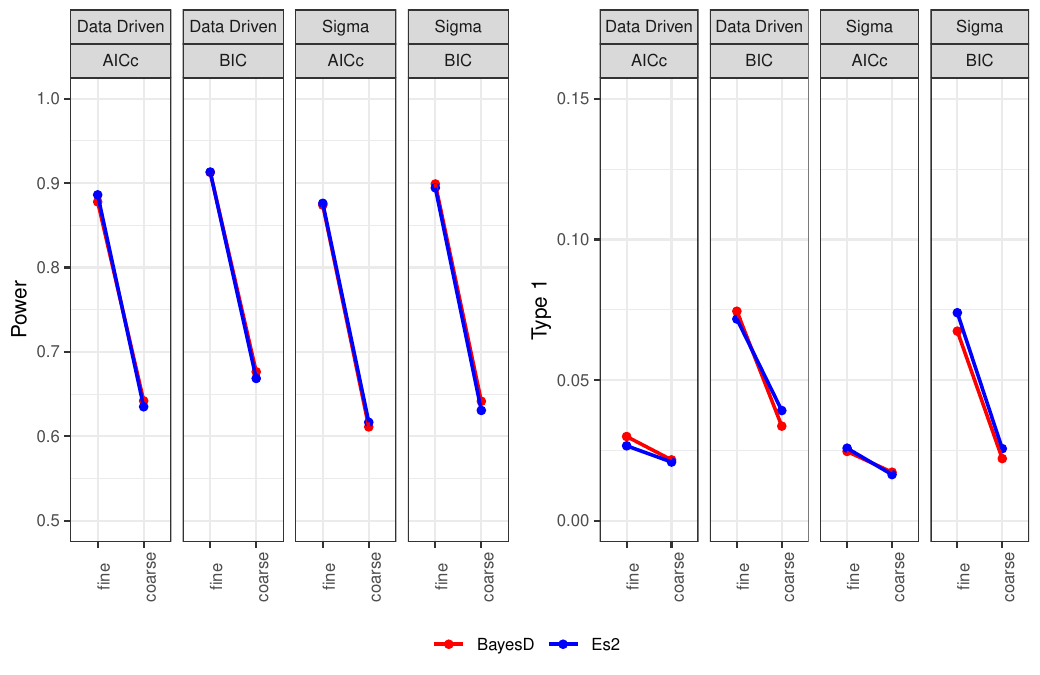}
     \caption{Comparison of (18, 22)  Bayesian D-optimal and balanced $E(s^2)$ SSD for 9 active effects with an effect magnitude of the following values: 10,8,5,3 and the rest 2 for unknown signs.}
    \label{fig:enter-label}
 \end{figure}

\section{Inconsistencies in Larger Screening Designs}

This section contains the remainder of the simulations from \cite{mee2017selecting} not shown in Section 2.2 of the main paper.  These are the complete simulation results for the $n=20$, $k=7$ designs.

\begin{figure}[H]
     \centering
     \includegraphics{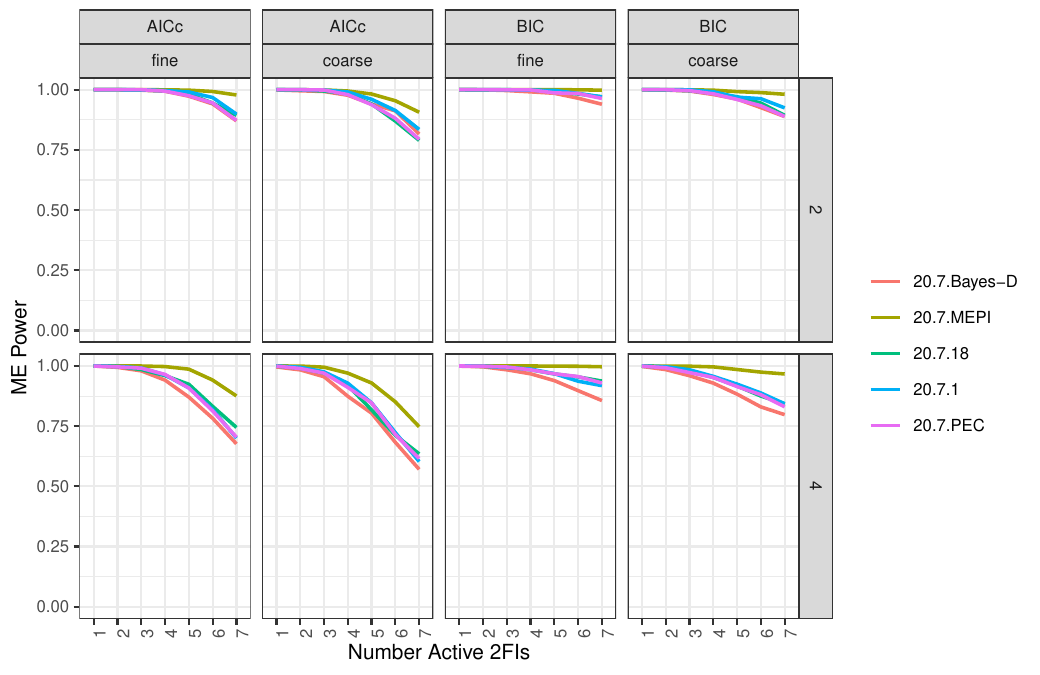}
     \caption{Power of the main effects for 2 or 4 (top row and bottom row, respectively) active two-factor interactions with true effects equivalent in magnitude to the main effects.}
    \label{fig:enter-label}
 \end{figure}

\begin{figure}[H]
     \centering
     \includegraphics{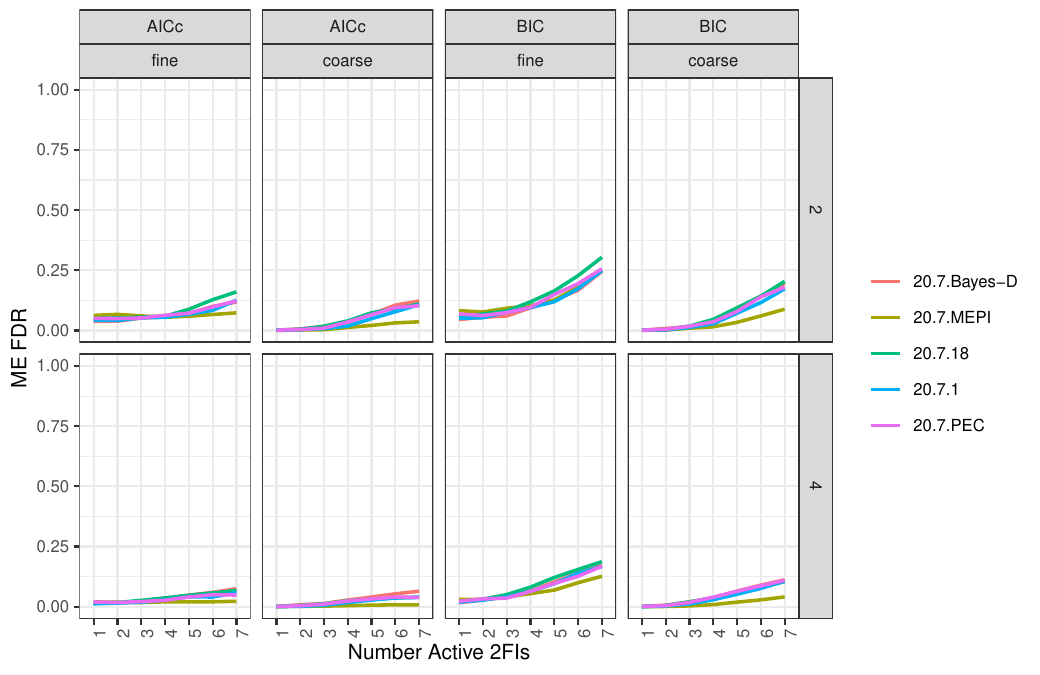}
     \caption{FDR of the main effects for 2 or 4 (top row and bottom row, respectively) active two-factor interactions with true effects equivalent in magnitude to the main effects. This figure corresponds to Figure 6 in the main document.}
    \label{fig:enter-label}
 \end{figure}

\begin{figure}[H]
     \centering
     \includegraphics{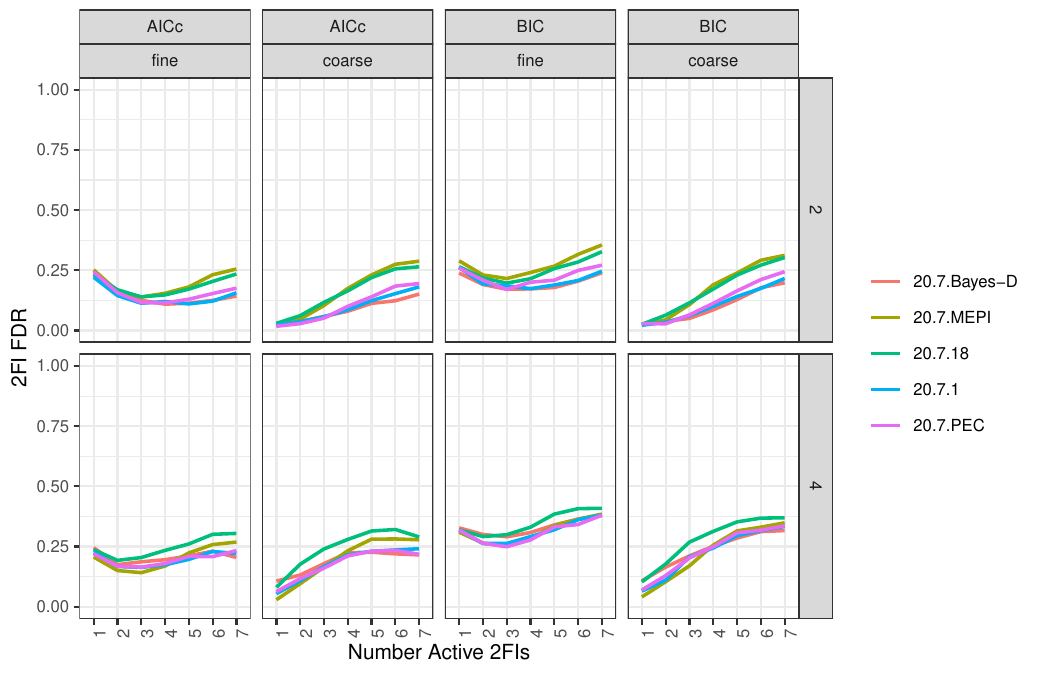}
     \caption{FDR of the two-factor interactions for 2 or 4 (top row and bottom row, respectively) active two-factor interactions with true effects equivalent in magnitude to the main effects. This figure corresponds to Figure 5 in the main document.}
    \label{fig:enter-label}
 \end{figure}

 \begin{figure}[H]
     \centering
     \includegraphics{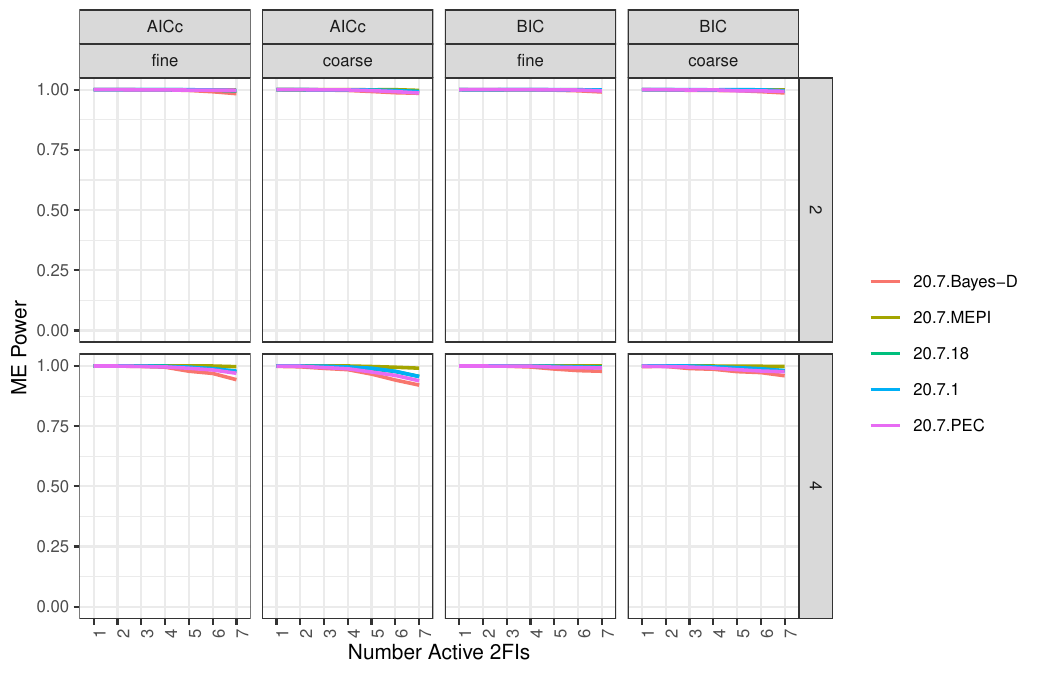}
     \caption{Power of the main effects for 2 or 4 (top row and bottom row, respectively) active two-factor interactions with true effects smaller in magnitude to the main effects.}
    \label{fig:enter-label}
 \end{figure}

 \begin{figure}[H]
     \centering
     \includegraphics{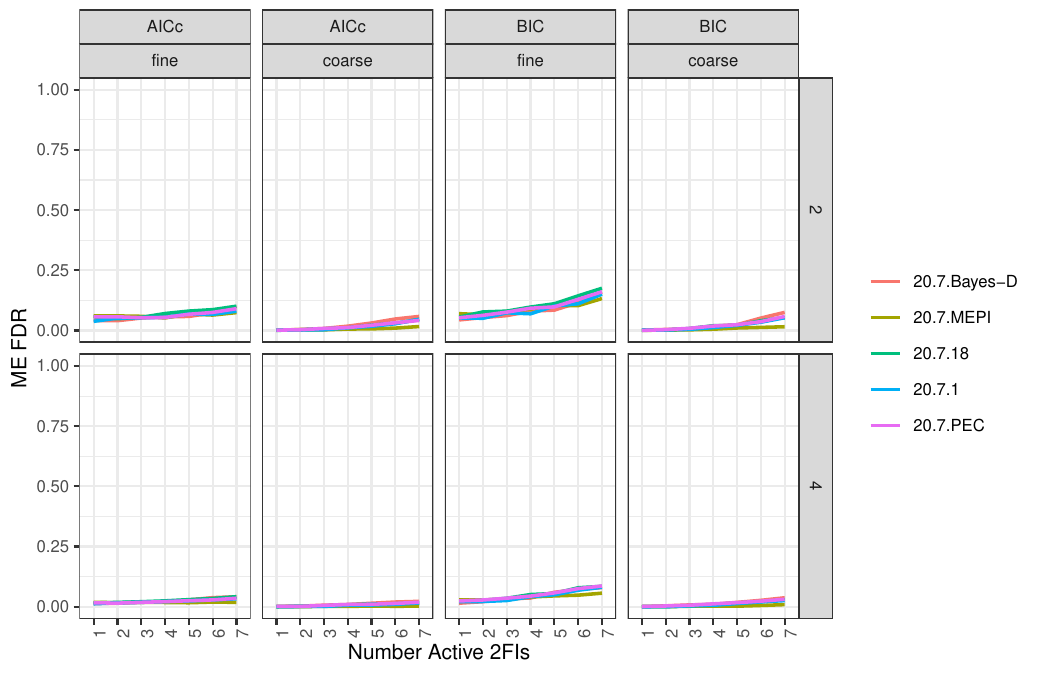}
     \caption{FDR of the main effects for 2 or 4 (top row and bottom row, respectively) active two-factor interactions with true effects smaller in magnitude to the main effects.}
    \label{fig:enter-label}
 \end{figure}

  \begin{figure}[H]
     \centering
     \includegraphics{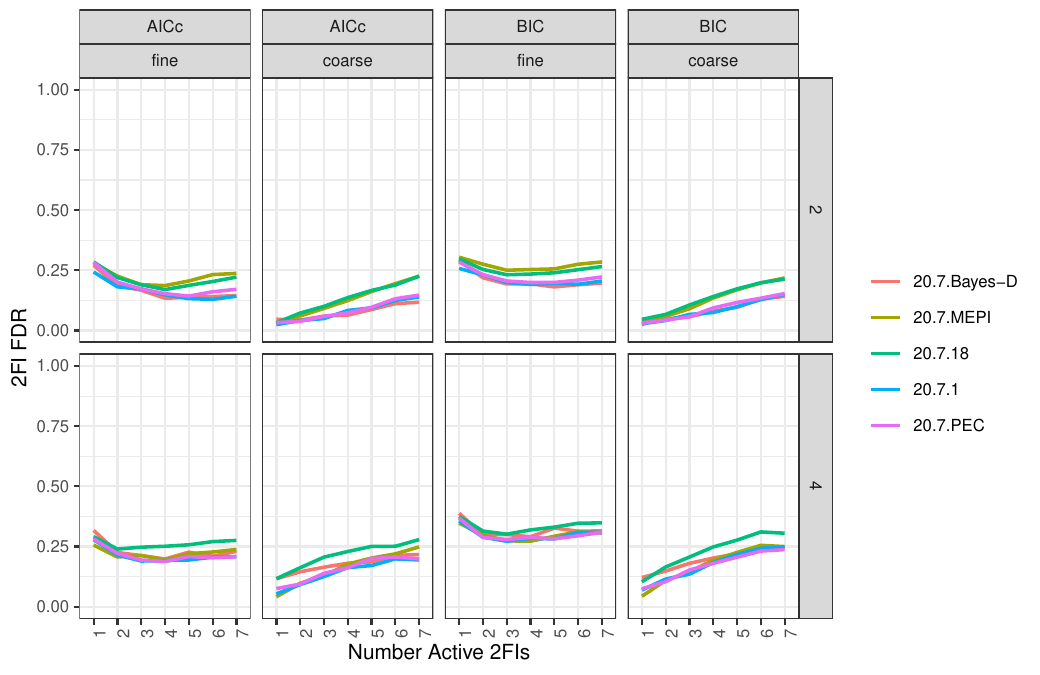}
     \caption{FDR of the two-factor interactions for 2 or 4 (top row and bottom row, respectively) active two-factor interactions with true effects smaller in magnitude to the main effects.}
    \label{fig:enter-label}
 \end{figure}

\section{Lasso Sign Recovery Probabilities}

Sign recovery at a fixed $\lambda$ involves two events of normal random variables, denoted $\mathcal{I}_\lambda$ and $\mathcal{S}_\lambda$. The $\mathcal{I}_\lambda$ event represents the event of estimating all inactive effects as 0.  The $\mathcal{S}_\lambda$ event represents the event of selecting all truly active factors and estimating them with the correct sign. Both  events must occur for sign recovery under the lasso. These two events are independent and therefore the probability of sign recovery under the lasso for a fixed $\X$ and $\betavec$ with sign vector $\zvec$ is defined as:
\begin{align}
\phi_\lambda(\X|\betavec)=P(\hat{\zvec}=\zvec|\X,\betavec)=P[\mathcal{I}_\lambda \cap \mathcal{S}_\lambda | \X, \betavec]=P[\mathcal{I}_\lambda | \X, \zvec] \times P[\mathcal{S}_\lambda|\X,\betavec],
\end{align}
\noindent where $\hat{\zvec}$ is the estimated sign vector. The computation of this probability requires knowledge of $\lambda$ and $\betavec$ which also implicitly requires knowledge of the set of active effects, $\mathcal{A}$. To handle these issues, \cite{stallrich2023optimal} evaluate $\phi_\lambda$ over all possible active sets of size $a$ (or, in some cases, a subset of all possible active sets) and sign vectors for a range of $\lambda$ values.  To make the computations even faster, they assumed the active effects had a constant absolute magnitude of $\beta$, being a user-specified minimum signal-to-noise ratio to detect. 

If the sign vector of the active effects is assumed to be known, then without loss of generality we may assume the signs are all positive and $\Phi_\lambda(\X | a, \beta)$ is defined to be the average of $\phi_\lambda(\X | \beta)$ over all possible active sets of size $a$. Alternatively, when one is unwilling to make the assumption of known effect signs, $\Phi_\lambda^{\pm}(\X | a, \beta)$ is calculated as the average of $\phi_\lambda(\X | \beta)$ over all possible active sets of size $a$ and all $2^a$ possible sign vectors for the active factors. \cite{stallrich2023optimal} do provide some results that halve the number of sign vectors ($2^{a-1})$) that need ot be considered in the computation.

\section{Comparing Larger Screening Designs via Sign Probabilities}

The graphs below show the sign recovery probabilities for the entire set of possible two-factor interactions from the scenarios presented in \cite{mee2017selecting}.
\newpage

\begin{figure}[H]
    \centering
   \includegraphics[width=6in]{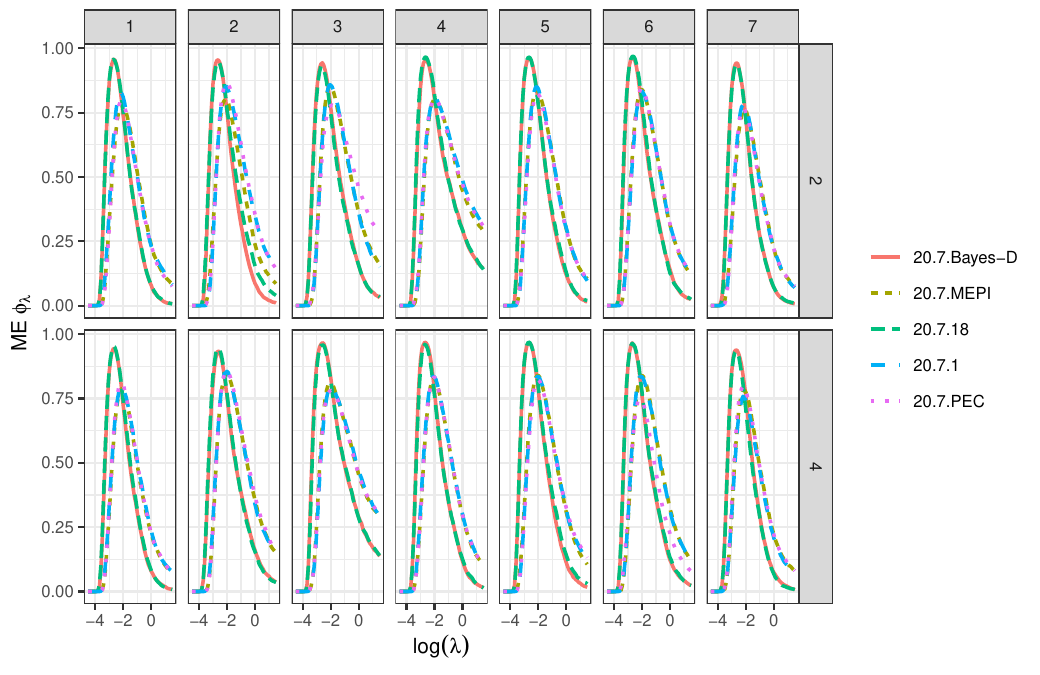}
   \caption{Comparison of $n=20$, $k=7$ designs using the simulated lasso sign recovery probability for the main effects only.  The sign recovery probability for main effects is compared for 2 and 4 active main effects and 1-7 active two-factor interactions (columns in the figure).}
   \label{fig:lasso_sim_ME_sup}
\end{figure}

\newpage

 \begin{figure}[H]
    \centering
   \includegraphics[width=6in]{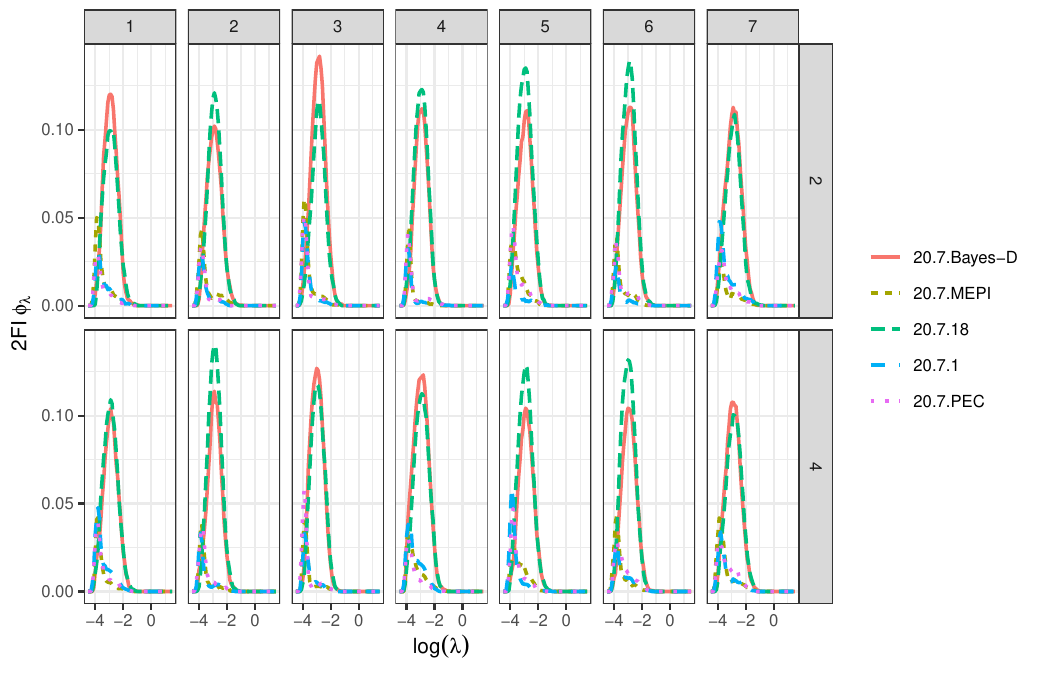}
   \caption{Comparison of $n=20$, $k=7$ designs using the simulated lasso sign recovery probability for the two factor interactions only.  The sign recovery probability for two factor interactions is compared for 2 and 4 active main effects (rows in the figure) and 1-7 active two-factor interactions (columns in the figure).}
   \label{fig:lasso_sim_2FI_sup}
\end{figure}
\end{document}